\documentclass[twocolumn]{aa}
\usepackage{graphicx}
\usepackage[varg]{txfonts}
\usepackage[colorlinks = true,
            linkcolor = blue,
            urlcolor  = blue,
            citecolor = blue,
            anchorcolor = blue]{hyperref}
\usepackage{bm}
\usepackage{soul}
\usepackage[dvipsnames]{xcolor}
\usepackage{verbatim}
\usepackage{mhchem}
\def\cii{[\ion{C}{II}] }
\def\hi{\ion{H}{I} }

\def\be{\begin{equation}}
\def\ee{\end{equation}}
\def\ba{\begin{eqnarray}}
\def\ea{\end{eqnarray}}
\def\eqi{\begin{equation}}
\def\eqf{\end{equation}}
\def\eqia{\begin{eqnarray}}
\def\eqfa{\end{eqnarray}}

\usepackage{ulem}
\usepackage{multirow}
\usepackage{booktabs}
\usepackage{relsize}
\usepackage{nameref}
\usepackage{array}
\usepackage{float}
\usepackage{stfloats}
\usepackage{amsmath}

\usepackage[nameinlink,noabbrev]{cleveref}
\Crefname{equation}{Eq.}{Eqs.}
\Crefname{eqnarray}{Eq.}{Eqs.}
\Crefname{section}{Sect.}{Sects.}
\Crefname{figure}{Fig.}{Figs.}
\crefname{equation}{Equation}{Equations}
\crefname{section}{Section}{Sections}
\crefname{figure}{Figure}{Figures}

\usepackage{enumitem}

\SetLabelAlign{Center}{\hfil#1\hfil}
\SetLabelAlign{CenterWithParen}{\hfil(\makebox[1.0em]{#1})\hfil}

\newcolumntype{C}[1]{>{\centering\arraybackslash}p{#1}}

\definecolor{mediumorchid}{RGB}{208,32,255}
\definecolor{forestgreen}{RGB}{0,179,0}

\newcommand{\der}{\mathrm{d}}

\newcommand{\Pwn}{P_\mathrm{WN}}

\newcommand{\rhombar}{\bar{\rho}_\mathrm{m}}

\newcommand{\Mpc}{\mathrm{Mpc}}
\newcommand{\keV}{\mathrm{keV}}

\usepackage{orcidlink}

\begin{document}

\title{Probing the warm dark matter mass with \cii intensity mapping}
\titlerunning{Probing the WDM mass with \cii intensity mapping}

\author{Elena~Marcuzzo$^{1,}$\thanks{\email{emarcuzzo@astro.uni-bonn.de}}\orcidlink{0009-0005-2491-8507}, Cristiano~Porciani$^{1,2,3,4}$\orcidlink{0000-0002-7797-2508}, Emilio~Romano-D\'iaz$^{1}$\orcidlink{0000-0002-0071-3217}, Azadeh Moradinezhad Dizgah$^{6}$\orcidlink{0000-0001-8841-9989}, Prachi~Khatri$^{8,1}$\orcidlink{0009-0009-1983-8333}, and Matteo Viel$^{2,3,4,5,7}$\orcidlink{0000-0002-2642-5707}}

\institute{$^{1}$ Argelander-Institut f\"ur Astronomie, Universit\"at Bonn, Auf dem H\"ugel 71, 53121 Bonn, Germany \\ $^{2}$ SISSA, International School for Advanced Studies, Via Bonomea 265, 34136 Trieste, Italy \\ $^{3}$ INAF, Dipartimento di Fisica -- Sezione di Astronomia, Universit\`a di Trieste, Via Tiepolo 11, 34131 Trieste, Italy \\ $^{4}$ IFPU, Institute for Fundamental Physics of the Universe, Via Beirut 2, 34151 Trieste, Italy \\ $^{5}$ INFN, National Institute for Nuclear Physics of the Universe, Via Valerio 2, 34127 Trieste, Italy \\ $^{6}$ Laboratoire d’Annecy de Physique Th\'eorique (LAPTh), CNRS/USMB, 99 Chemin de Bellevue BP110, Annecy -- F-74941 -- France \\ $^{7}$ ICSC -- Centro Nazionale di Ricerca in High Performance Computing, Big Data and Quantum Computing, Via Magnanelli 2, Bologna, Italy \\ $^8$ Cardiff Hub for Astrophysics Research and Technology, School of Physics and Astronomy, Cardiff University, Queen’s Buildings, Cardiff CF24 3AA, UK}

\date{Received 8 December 2025 / Accepted 12 January 2026}
\authorrunning{E. Marcuzzo et al.}

\abstract
{The nature of dark matter (DM) is still debated. While cold DM (CDM) is the standard paradigm, warm DM (WDM) composed of thermal relics may ease some small-scale tensions in the $\Lambda$CDM framework. Line-intensity mapping (LIM) offers a novel probe of DM properties.}  
{To explore the potential of LIM surveys in constraining the WDM particle mass ($m_\mathrm{WDM}$) by means of the \cii power spectrum (PS), we provide forecasts for the Deep Spectroscopic Survey (DSS) to be performed with the Fred Young Submillimeter Telescope at $z\simeq3.6$ and extend the analysis to larger sky coverage, higher sensitivity, and/or increased spectral resolution.}  
{We developed a formulation for the \cii PS based on the halo-model approach, incorporating the uncertainty in the luminosity function (LF) through two alternative parameterisations, one optimistic and the other more conservative. We performed a Bayesian analysis on mock data to derive constraints on $m_\mathrm{WDM}$.}  
{In a CDM universe, the DSS yields lower limits on $m_\mathrm{WDM}$, at a $95\%$ credibility level, of $1.10$ keV and $0.58$ keV when considering the optimistic and pessimistic LF ($\alpha = -1.1$), respectively.
Ambitious surveys can improve these figures to $5.82$ keV and $1.90$ keV, and assuming a steeper faint-end slope ($\alpha = -1.9$) further boosts the constraints even beyond those obtained in the optimistic scenario. A fivefold increase in spectral resolution enhances sensitivity to the damping scale associated with redshift-space distortions, tightening the constraints on $m_\mathrm{WDM}$ by a factor of up to $\sim1.8$. Finally, Bayesian inference on mock data with $m_\mathrm{WDM}=3 \ \mathrm{keV}$ results in a well-constrained and unbiased posterior only in futuristic survey setups.}
{Upcoming LIM surveys can provide meaningful limits on $m_\mathrm{WDM}$, although the negligible contribution from small haloes reduces the constraining power of the \cii PS. Future progress will benefit from combining multiple redshifts and emission lines, opening the way to competitive constraints on the nature of DM.}

\keywords{methods: statistical -- galaxies: high-redshift -- galaxies: luminosity function, mass function -- dark matter -- \\ large-scale structure of Universe
}

\maketitle
\section{Introduction}\label{sec:intro}
Over the past decade, line-intensity mapping (LIM) has gained attention as a promising strategy to probe both the large-scale distribution of matter and the galactic properties across cosmic time \citep[see][for a general overview]{Kovetz+17,Bernal22_review}. While the 21 cm line remains the most extensively studied tracer \citep[see e.g.][]{Morales_Wyithe_10,Pritchard_Loeb12,Liu_Shaw_20}, the scope of LIM studies has rapidly broadened to include several other emission lines, with multiple pathfinder surveys now operating or planned across a wide range of wavelengths.\footnote{\url{https://lambda.gsfc.nasa.gov/product/expt/lim_experiments.html}}

Alongside this experimental progress, theoretical work has explored how LIM observables relate to the physics of reionisation and astrophysical properties \citep[see e.g.][]{Lidz+09,Gong+12,Visbal+15,Comaschi-Ferrara_16,Breysse-Rahman_17,Breysse-Alexandroff_19,Sato-Polito+20,Zhou+21}. For instance, the amplitude and shape of the LIM power spectrum (PS) encode information about the abundance, clustering, and luminosity distribution of the emitting galaxies. Therefore, the PS itself, in combination with other statistics, provides a means to constrain both the moments and parametric form of the luminosity function (LF), thus offering a new route to quantify galaxy properties in the early Universe \citep[see e.g.][]{Li+16,Ihle+19,Marcuzzo+25}.

Beyond astrophysical applications, LIM is increasingly being recognised as a tool with significant potential for cosmological tests. Because the large-scale PS traces the underlying matter distribution---modulated by biasing, redshift-space distortions (RSDs), and emission properties---it can be used to extract cosmological information \citep[see e.g.][]{Karkare+18,Azadeh18,Bernal+19,Azadeh19,Munoz+20,Bauer+21,Bernal+21,Azadeh22b,Azadeh22a, Karkare22, Adi23, Azadeh23,Fronenberg24,Shmueli25}, similar to galaxy clustering or the Lyman-$\alpha$ forest, but it extends to fainter populations and less explored epochs. This makes LIM particularly appealing for probing non-standard cosmological scenarios \citep{snowmass21}, despite the additional challenges posed by line interlopers, bright continuum foregrounds, and uncertainties in modelling the signal that remain more severe than in mature observational probes \citep[see e.g.][and references therein]{Schaan-White_21a,Schaan-White_21b}.

Continuing the endeavour of the community in identifying possible applications of LIM in cosmological contexts, we investigate its potential to shed new light on the nature of dark matter (DM), one of the biggest unsolved problems in physics (see \citealp{Cirelli_DMRev} for a recent review). In the standard model of cosmology, DM is assumed to be `cold', meaning that the free-streaming length associated with its thermal velocity in the matter-dominated era is negligibly small compared to the linear size of the perturbations that lead to the formation of dwarf galaxies. Popular particle candidates for cold DM (CDM) include axions \citep{Weinberg_78,Preskill+83}---which are non-thermally produced in the early Universe---and thermal relics \citep[e.g.][]{Hochberg+14,Arcadi+18}, which decouple from thermal equilibrium with standard-model particles and freeze out during the radiation-dominated era. CDM relics are non-relativistic when they decouple from the primordial plasma and are thus trapped within the primordial density perturbations produced during inflation.

Over time, some tensions have emerged between observations on sub-galactic scales and the predictions of the CDM scenario based on N-body simulations  \citep[e.g.][and references therein]{Bullock-Boylan-Kolchin_17}. Although baryonic physics could mitigate the discrepancies, it is possible to devise a scenario in which the DM particles are relativistic at decoupling and become non-relativistic during the radiation-dominated era. This causes the particles to freely stream out of kiloparsec-scale potential wells, giving rise to a cutoff in the linear matter PS relative to CDM \citep[e.g.][]{Bond-Szalay-83} and resulting in a deficit of small haloes and delayed galaxy formation in low-mass systems---features that can be probed observationally.
This `warm' DM \citep[WDM; e.g.][]{Bode+01,Viel05_WDM} scenario can be tested against observations using the flux PS of the Lyman-$\alpha$ forest \citep{Viel+13, Irsic+17, Villasenor+23, Irsic+24}, fluctuations in the density of cold stellar streams \citep{Banik+21}, dwarf-galaxy counts \citep{Nadler+21}, the gravitational lensing of distant quasars \citep{Gilman+20}, and combinations of these strategies \citep{Zelko+22}.
Current data put a lower bound on the mass of the WDM thermal relics of a few kiloelectronvolts---at a 95\% credibility level (CL).

A probe that is directly sensitive to the abundance of low-mass haloes is therefore particularly valuable since WDM primarily modifies structure formation below a well-defined mass scale. LIM is a natural candidate for these studies. Previous work has explored the impact of WDM on the LIM signal of the 21 cm line during \citep{Sitwell+14} and after \citep{Carucci15_WDM_HI} reionisation. Somewhat counterintuitively, \cite{Carucci15_WDM_HI} found enhanced 21 cm fluctuations in WDM relative to CDM because the depletion of low-mass haloes, combined with observational constraints on the cosmic abundance of neutral hydrogen, shifts more \ion{H}{I} into massive, highly biased haloes. However, they also concluded that SKA1-LOW would be unable to distinguish WDM from CDM at $z=5$ without several thousand hours of integration. This limitation reflects astrophysical considerations: In the post-reionisation Universe, diffuse gas in haloes with $M\lesssim 10^9\,\mathrm{M_\odot}$ is expected to photoevaporate rapidly due to insufficient self-shielding against the ultraviolet background \citep[e.g.][]{Rees_86,Barkana-Loeb-99}. For WDM particle masses consistent with current limits, the associated cutoff in the halo mass function (HMF) lies close to this evaporation scale, implying that little 21 cm emission originates from the low-mass CDM haloes absent in WDM.

These considerations motivate the exploration of tracers of star-formation activity, such as \cii\!, 
CO rotational transitions, and Balmer recombination emission. 
In particular, \cii and CO trace cold, dense gas associated with the inner star-forming regions of galaxies---gas that is far less susceptible to photoevaporation and therefore provides a cleaner probe of the low-mass structures where WDM and CDM predictions diverge most. This expectation is supported by both theoretical and observational studies showing that reionisation primarily removes diffuse, low-density gas from dwarf-galaxy haloes while leaving the dense, self-shielded component largely intact. As a result, star formation in low-mass systems is expected to taper gradually rather than cease abruptly \citep{SusaUmemura2004,Hoeft2006,Okamoto2009,Sawala2012,Simpson2013}. Recent simulations further reveal that the survival and re-ignition of star formation depend sensitively on the retention of cold, dense gas and on halo concentration, with some low-mass galaxies capable of initiating or renewing star formation as late as $z\simeq 2$ in favourable conditions \citep{Moreno2025}.

Observational evidence echoes this diversity. Nearby dwarfs display a wide range of star-formation histories shaped by reionisation, including systems that quenched early, systems that experienced delayed ignition, and `two-component' galaxies where old and young populations overlap \citep{BenitezLlambay2015}. Deep JWST imaging of isolated low-mass galaxies, such as Leo~P, reveals early star formation followed by a long pause after reionisation and a subsequent reignition at later times \citep{McQuinn2024}. Complementary HST and VLA studies of ultra-faint dwarfs near M31 show similarly varied behaviour: Some quenched shortly after reionisation, while others formed stars well into the post-reionisation era despite extremely low present-day gas content \citep{Jones2025}. Even in isolated systems, spatially resolved JWST data reveal radial and azimuthal variations in stellar age gradients consistent with localised star-formation episodes triggered by the interaction of dense gas with the surrounding medium \citep{Cohen2025}.

Taken together, these results emphasise that dense star-forming gas can persist in low-mass haloes long after reionisation, making tracers such as \cii ideal for probing the small-scale structure most affected by WDM. In this work, we investigate whether future ground-based LIM experiments targeting the \cii fine-structure line can provide competitive constraints on the WDM particle mass. The \cii transition arises mainly in photon-dominated regions at the boundaries of molecular clouds exposed to far-ultraviolet radiation, making it an excellent probe of ongoing star formation. Although previous estimates suggest that the \cii LIM signal at $z\simeq 5$ is dominated by haloes of $10^{11}$--$10^{12} \ \mathrm{M_\odot}$ \citep{yue+15}, contributions from less massive galaxies may remain non-negligible, particularly at lower redshifts. This makes the \cii LIM PS a promising observable for constraining the abundance of the low-mass haloes suppressed in WDM cosmologies.

To explore the potential of this approach, we adopted a representative observational setup based on the Deep Spectroscopic Survey (DSS), to be carried out with the 6 m Fred Young Submillimeter Telescope (FYST) \citep{CCATprime_collab}. We selected this ground-based survey, already scheduled for deployment, as it offers a practical near-term route to carrying out LIM observations. Thanks to its high and dry location, which enables efficient sub-millimetre measurements, combined with a wide field of view and fast mapping speed, this setup provides both the sensitivity and the survey volume required for a robust PS measurement.
In addition, the DSS is planned to target two fields that overlap with the Extended Cosmic Evolution Survey (E-COSMOS; \citealp{Aihara_ecosmos_18}; see also \citealp{Scoville_cosmos_07}), the Extended Chandra Deep
Field South \citep[E-CDFS;][]{Lehmer+05}, and the Hubble Ultra Deep Field \citep[H-UDF;][]{Beckwith+06} survey areas, providing an almost unique opportunity to obtain a comprehensive multi-wavelength characterisation of the same underlying galaxy population. We assess how survey design choices---such as sky coverage, instrumental sensitivity, and spectral resolution---affect the constraining power of the \cii LIM signal. Focusing on a redshift window centred at $z \simeq 3.6$, we evaluate the extent to which the \cii PS can reveal the small-scale suppression that characterises WDM models, offering a novel and complementary path to probing the nature of DM beyond the CDM paradigm.

To enable comparison with other results from the literature, we also express our results in terms of fuzzy DM \citep[FDM; e.g.][]{Irsic+17_fdm,Lazare+24}, an alternative model in which DM consists of ultra-light bosons that inhibit structure formation on small scales due to quantum effects. This conversion is possible because both WDM and FDM produce a characteristic cutoff in the linear matter PS, albeit for different physical reasons. By matching the half-mode wavenumbers (i.e. the scale at which this suppression occurs), we can translate a constraint on the WDM particle mass into an effective constraint on the FDM boson mass. We note, however, that this correspondence is approximate and model dependent.

The paper is organised as follows.
Sect.~\ref{sec:WDM_FDM} introduces WDM and FDM scenarios, detailing how the beyond-CDM PS is derived from its CDM counterpart.
In Sect.~\ref{sec:halo_model} we briefly summarise the halo-model formalism used to compute the LIM PS.
In Sect.~\ref{sec:CII_LF_HAM} we consider the high-redshift \cii LF and describe the abundance-matching procedure used to connect halo masses and \cii\!-line luminosities.
Sect.~\ref{sec:CII_power_spectrum} introduces the different survey setups considered throughout this work and presents the predicted \cii PS with relative uncertainties.
In Sect.~\ref{sec:bayesian_inference} we outline our Bayesian inference pipeline applied to mock data and report the resulting constraints on the WDM particle mass. We show their conversion to FDM as well.
In Sect.~\ref{sec:discussion}, we discuss critical aspects of our modelling and inference strategy, with particular attention given to the relative contribution of different halo masses to the LIM signal and, consequently, to the impact of varying the faint-end slope of the \cii LF.
Finally, in Sect.~\ref{sec:summary_conclusions} we provide a summary of our findings and draw the main conclusions. The possibility of including other redshift bins and/or emission lines is also taken into account.

We assumed a flat Friedmann-Lema\^{i}tre-Robertson-Walker (FLRW) cosmology, adopting $h=0.674$ for the dimensionless Hubble parameter. The present-day density parameters are set to $\Omega_\mathrm{m}=0.315$ for matter, $\Omega_\mathrm{b}=0.049$ for baryons, and $\Omega_\Lambda=0.685$ for the cosmological constant. The primordial PS of density fluctuations is described by a spectral index $n_\mathrm{s}=0.965$ and a normalisation $\sigma_8=0.811$. The linear matter PS in the CDM scenario, $P_\mathrm{CDM}$, is obtained using the Code for Anisotropies in the Microwave Background \citep[\texttt{CAMB};\footnote{\href{https://camb.info/}{https://camb.info/}}][]{Lewis+00}.

\section{Cold, warm, and fuzzy dark matter}
\label{sec:WDM_FDM}

Fermionic DM of mass $m_\mathrm{WDM}\sim$ keV$/c^2$  produced in thermal equilibrium results in suppressed small-scale clustering with respect to CDM.
This is because the DM particles are relativistic at production and free stream out of primordial perturbations within a mass-dependent characteristic scale.  
We parameterised the damping in the linear matter PS as
\begin{equation}
    P_\mathrm{WDM}(k) = T^2(k) \, P_\mathrm{CDM}(k) \; , 
    \label{eq:P_WDM}
\end{equation}
where \citep[e.g.][]{Viel05_WDM}
\begin{equation}
    T(k) = \left[1+\left(\tau k\right)^{2\nu}\right]^{-5/\nu} \; ,
    \label{eq:WDMtransfer}
\end{equation}
with $\nu = 1.12$ and 
\begin{equation}
    \tau = 0.049 \ \left(\frac{m_\mathrm{WDM}}{1\, \keV}\right)^{-1.11} \ \left(\frac{\Omega_\mathrm{m}}{0.25}\right)^{0.11} \ \left(\frac{h}{0.7}\right)^{1.22} \ h^{-1}\, \Mpc \;.
\end{equation}
This was obtained by fitting the output of Boltzmann codes for wavenumbers $k<5 \, h \, \Mpc^{-1}$.

For both CDM and WDM models, we computed the HMF and linear bias by using the fits to numerical simulations \citep[][]{ST_99,ST_01}.
Following \cite{parimbelli_21}, we determined the linear mass variance smoothed on the halo scale with the so-called smooth-$k$ window function, which avoids the excess of low-mass haloes predicted by the standard top-hat filter in models with a small-scale cutoff while providing a reasonable fit to N-body simulation results:
\begin{equation}
    \widetilde{W}(k) = \left[1+(kR_\mathrm{s})^{\beta_\mathrm{s}}\right]^{-1} \; ,
\end{equation}
with $\beta_\mathrm{s}=4.8$ and $R_\mathrm{s}=3.3^{-1}\,\sqrt[3]{3M/(4 \pi \rhombar)}$, where $M$ denotes the halo mass and $\rhombar$ is the mean comoving matter density.
For completeness, we include in Appendix~\ref{app:HMF_bias} a graphical illustration of the effects of WDM relative to CDM on the PS, HMF, and halo bias.

We also considered the possibility that DM is made of ultra-light bosons  with kiloparsec-scale De Broglie wavelengths, the FDM.
In this case, we determined the mass of the FDM particles, $m_\mathrm{FDM}$, through the relation \citep{conversion_fuzzyDM}
\begin{equation}
    k_{0.5}=4.5 \,\left(\frac{m_\mathrm{FDM}\,c^2}{10^{-22} \mathrm{eV}}\right) ^{4/9} \ \mathrm{Mpc}^{-1}\;,
    \label{eq:FDM_constraints}
\end{equation}
with $T^2(k_{0.5})=0.5$.

\section{The theoretical \cii power spectrum}\label{sec:halo_model}

We describe the large-scale \cii intensity fluctuations using the halo-model formalism, which provides a physically motivated framework for linking the statistical properties of matter to those of the underlying halo population \citep{Cooray_Sheth_02}. This approach has been extensively applied to LIM, where the line emission is connected to haloes through empirically or theoretically motivated luminosity--mass relations \citep[e.g.][]{Silva+15,Gong+17}.
A summary of the main aspects of our implementation is presented below, while further details and related prescriptions can be found in \citet[][M25 hereafter]{Marcuzzo+25}.

Under the assumption that line emission occurs in DM haloes and neglecting absorption and scattering, the mean specific intensity observed at redshift $z$ for a transition with rest-frame frequency $\nu_{\rm rf}$ is given by
\begin{equation}
    \bar{I}_\nu(z) = \frac{c}{4\pi H(z)\, \nu_{\rm rf}}\, \bar{\rho}_\mathrm{L}(z) \ ,
    \label{eq:mean_intensity}
\end{equation}
where $\bar{\rho}_\mathrm{L}(z)$ is the mean comoving luminosity density, i.e. the first moment of the LF.

To relate galaxy luminosities to halo properties, we introduce the conditional luminosity function (CLF), $\phi(L|M,z)$, which quantifies the average number of galaxies with luminosity $L$ inside haloes of mass $M$ at redshift $z$. This allows one to express the LF as
\begin{equation}
    \Phi(L,z) = \int_0^\infty \phi(L|M,z)\, \frac{\der  \bar{n}_\mathrm{h}}{\der M}(M,z)\, \der M \ .
\end{equation}
We further define the $n$-th moments of the CLF as
\begin{equation}
    \eta_n(M,z) = \int_0^\infty L^n\, \phi(L|M,z)\, \der L \ ,
    \label{eq:LF_moments}
\end{equation}
which correspond to the total $L^n$-weighted luminosity in haloes of mass $M$. For instance, $\eta_1$ gives the mean total \cii luminosity at a fixed halo mass, and thus the mean emissivity becomes
\begin{equation}
    \bar{\rho}_L(z) = \int_0^\infty \eta_1(M,z)\, \frac{\der  \bar{n}_\mathrm{h}}{\der M}(M,z)\, \der M \ .
\end{equation}
These moments capture the distribution of line luminosities within haloes of varying mass and play a central role in the following determination of the LIM PS model.

Under the assumptions of linear biasing, Poisson statistics, and the distant-observer approximation, the redshift-space PS of intensity fluctuations can be decomposed as
\begin{equation}
    P(k, \mu, z) = P_{\rm clust}(k, \mu, z) + P_{\rm shot}(z) \ ,
    \label{eq:theoretical_LIM_PS}
\end{equation}
where $P_{\rm clust}$ describes the large-scale clustering and $P_{\rm shot}$ accounts for the shot noise due to the discrete nature of sources. Moreover, $\mu = \hat{\mathbf{k}} \cdot \hat{\mathbf{n}}$ is the cosine of the angle between the wavevector and the line of sight.

On large scales, the clustering term is sourced by biased, linearly evolving tracers of the matter field and can be modelled as follows:
\begin{equation}
    P_{\rm clust}(k, \mu, z) = \bar{I}_\nu^2(z) \left[ b(z) + f(z)\mu^2 \right]^2 \mathcal{D}(k, \mu, z) \ P_{\rm m}(k, z) \ ,
\end{equation}
where $b(z)$ denotes the luminosity-weighted linear bias, $f(z)$ the linear growth rate, and $P_{\rm m}(k,z)$ the linear matter PS, taken as $P_\mathrm{CDM}$ or $P_\mathrm{WDM}$ depending on the underlying cosmological model. The damping function \(\mathcal{D}\), which models RSDs, can be parameterised using various functional forms. In this work, we adopt the squared Lorentzian form,
\begin{equation}
\mathcal{D}(k,\mu,z) = \left[1 + \frac{(k \mu \sigma)^2}{2} \right]^{-2} \ ,
\end{equation}
although our conclusions do not significantly change if a different commonly used shape is employed. The parameter $\sigma$ indicates a characteristic comoving displacement that encodes the strength of the damping generated by the RSDs, which is expected to be roughly comparable to the pairwise velocity dispersion scaled by $(aH)^{-1}$.
Finally, the luminosity-weighted bias, $b(z)$, can be expressed in terms of $\bar{\rho}_\mathrm{L}$ as
\begin{equation}
    b(z) = \frac{1}{\bar{\rho}_L(z)} \int_0^\infty \eta_1(M,z)\, b_h(M,z)\, \frac{\der \bar{n}_h}{\der M}(M,z)\, \der M.
\end{equation}

On the other hand, the shot-noise component is given by
\begin{equation}
    P_{\rm shot}(z) = \frac{\bar{I}_\nu^2(z)}{\bar{n}_{\rm eff}(z)} \ ,
\end{equation}
where $\bar{n}_{\rm eff}$ is the effective number density of emitters and can be defined as
\begin{equation}
    \bar{n}_{\rm eff}(z) = \left[ \frac{1}{\bar{\rho}_L^2(z)} \int_0^\infty \eta_2(M, z) \frac{d\bar{n}_h}{dM}(M,z)\,dM \right]^{-1} \ .
\end{equation}

\section{Mass--luminosity relation for \cii emitters}\label{sec:CII_LF_HAM}

Building on the halo--galaxy connection for \cii emitters identified in the \textsc{Marigold} simulations \citep{Khatri+24_marigold}, and following the methodology of M25, we assigned a \cii luminosity, $L$, to each halo via abundance matching, assuming $L$ increases monotonically with halo mass.
This procedure implicitly assumes that \cii emission is dominated by a single source per halo and neglects scatter in luminosity at fixed mass.

To implement this procedure, one requires the LF of \cii emitters at the relevant redshift. Despite recent observational progress, this function remains highly uncertain owing to the limited number of detections currently available.
We therefore relied on model predictions calibrated against existing measurements to construct a plausible LF, which serves as the basis for the abundance-matching relation.

We modelled the \cii LF using the Schechter form
\begin{equation}
   \Psi(L) \equiv \frac{\mathrm{d}n}{\mathrm{d}\log_{10} L}
   = \Psi_* \left(\frac{L}{L_*}\right)^{1+\alpha}
     \exp\!\left(-\frac{L}{L_*}\right) \; ,
   \label{schechter_function_psi}
\end{equation}
where $\Psi_*$, $L_*$, and $\alpha$ denote the normalisation, characteristic luminosity, 
and faint-end slope, respectively.  
To bracket the current observational uncertainties, we adopted two parallel LF parameterisations (see Fig.~1 in M25).  
The first is a high-normalisation, or `optimistic', model based on the multi-wavelength and multi-redshift 
fit of \citet{yan_20}.  
The second is a low-normalisation, or `pessimistic',\footnote{We refer to this model as `pessimistic' 
because it is derived from a targeted survey (ALPINE), where incomplete sampling of the galaxy population 
may lead to an underestimation of the true LF.} model obtained by fitting only the targeted sample from the ALPINE 
\cii survey \citep{lefevre_20,bethermin_20,faisst_20}.  
For the optimistic and pessimistic cases, we used $(\Psi_*, L_*, \alpha) = (10^{-3.08}\,\mathrm{Mpc}^{-3}\textrm{dex}^{-1}, 10^{9.5}\,L_\odot, -1.1)$  and
 $(10^{-3.02}\,\mathrm{Mpc}^{-3}\textrm{dex}^{-1}, 10^{8.73}\,L_\odot, -1.1)$, respectively.
These two parameterisations define the fiducial models analysed throughout this work.  
To assess the sensitivity of our results to the assumed faint-end shape, we additionally performed a series of 
fits to the ALPINE data with fixed slopes in the range $\alpha \in [-1.9,-0.5]$.

\begin{figure*}
\centering
\includegraphics[width=0.39\textwidth]{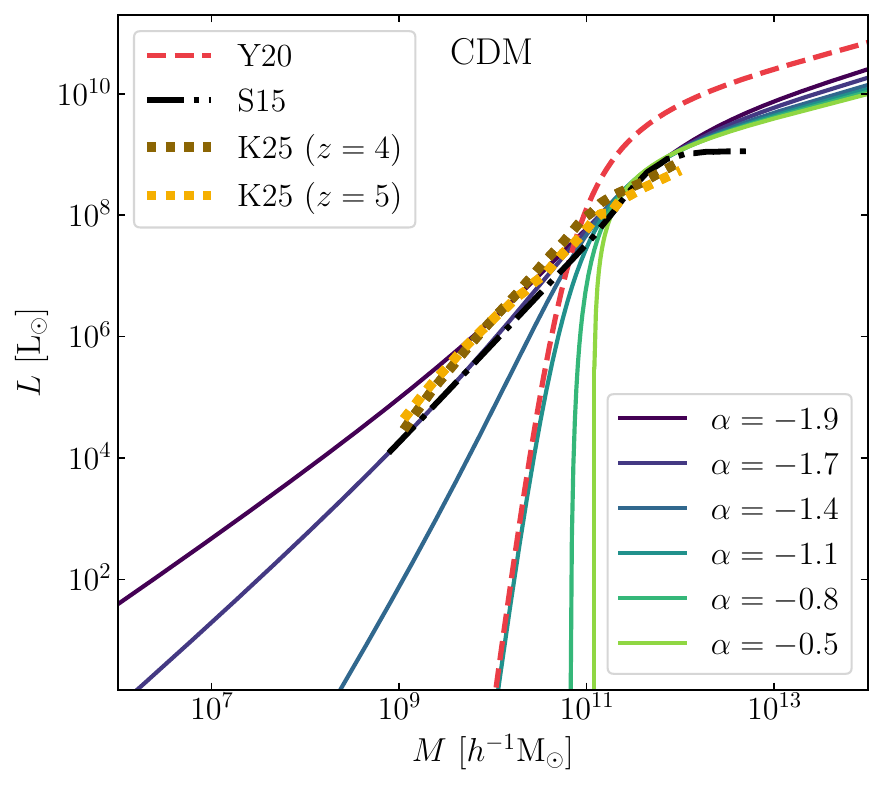}  \hspace{1.5cm}
\includegraphics[width=0.39\textwidth]{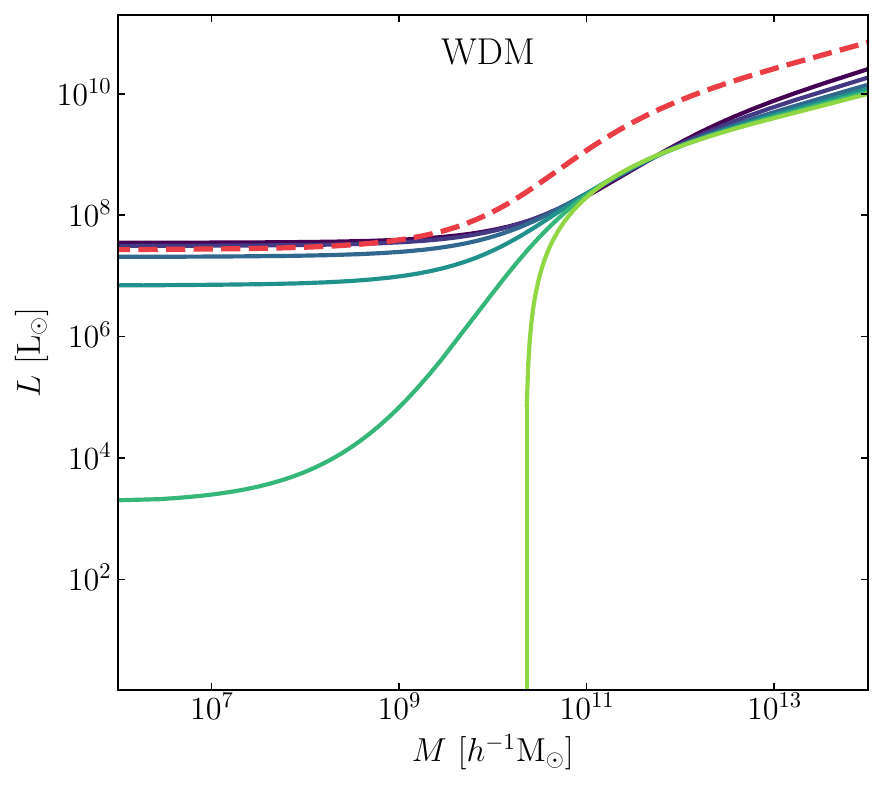} 
\caption{\cii luminosity--halo mass relation inferred from abundance matching. In both panels, the solid lines correspond to the $L(M)$ relations derived from our Schechter fits to the ALPINE data with fixed values for $\alpha$. The dashed red line is based on the fit by \citet[][Y20]{yan_20}. The left and right panels assume a CDM and a $0.5\ \mathrm{keV}$ HMF, respectively (see Appendix~\ref{app:HMF_bias} for a direct comparison between the two scenarios).
In the left panel, for comparison, the dot-dashed black line indicates the $L(M)$ relation obtained by \citet[][S15]{Silva+15}, while the dotted gold and dark gold lines show results from the \textsc{Marigold} simulations \citep[][K25]{Khatri+24_marigold} at $z=5$ and $z=4$, respectively. }
\label{fig:HAM}
\end{figure*}

\begin{figure}
\centering
\includegraphics[width=0.39\textwidth]{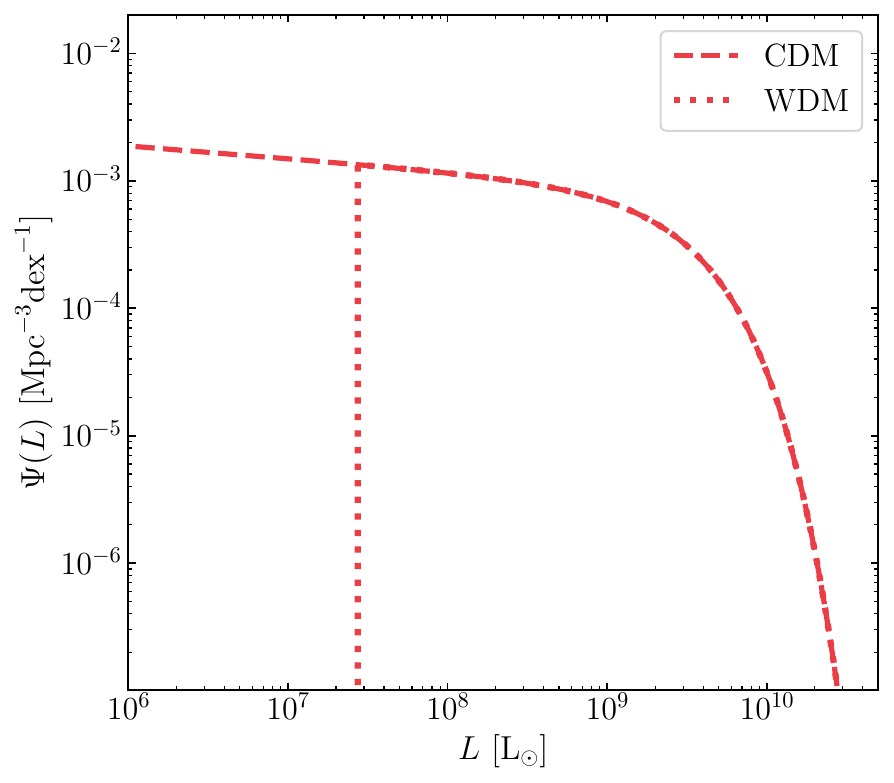} \caption{Reconstructed \cii LFs obtained by combining the abundance-matched 
$L(M)$ relations with the HMFs of the CDM and WDM cosmologies. The figure uses the LF of our optimistic case and adopts a WDM particle mass of $0.5\ \mathrm{keV}$. While the CDM model reproduces the input LF across all luminosities, the WDM case shows a sharp downturn at the faint end, reflecting the dearth of low-mass haloes and the corresponding minimum luminosity imposed by the abundance-matching procedure.}
\label{fig:reconstructed_LF}
\end{figure}

In our abundance-matching procedure, the observed \cii LF was used as a constraint that both cosmological scenarios must satisfy. Since the LF is an empirical property of the galaxy population, its bright end must be reproduced in both CDM and WDM. Only the faint end is allowed to differ between the two models, as it is not yet well constrained observationally and naturally reflects the differing halo abundances predicted by each cosmology.
The resulting abundance-matching relations are shown in Fig.~\ref{fig:HAM} for the two DM scenarios (left: CDM; right: WDM). The WDM case adopts an extreme particle mass of $0.5\,\mathrm{keV}$ to maximise the contrast with CDM. Unless the faint-end slope of the LF is very shallow, the scarcity of low-mass haloes in WDM implies the existence of a minimum \cii luminosity: There are simply not enough haloes to host galaxies below this threshold. As a result, the abundance matching cannot be fully satisfied, and the input LF must be truncated in the WDM case (see Fig.~\ref{fig:reconstructed_LF}). This minimum luminosity corresponds to the horizontal asymptotes of the $L(M)$ relation at low halo masses.
This feature partly reflects our assumption of one source per halo, but it arises primarily because the WDM HMF turns over and rapidly declines at low masses (see Appendix~\ref{app:HMF_bias}). The resulting plateau in the $L(M)$ relation is therefore an artefact of the abundance-matching procedure, signalling the drop in halo abundance rather than a physical property of the luminosity--mass relation. In the halo model of Sect.~\ref{sec:halo_model}, the effective PS components depend on the product $L(M)\,\mathrm{d}\bar{n}_\mathrm{h}/\mathrm{d}M$, so the fall-off of the HMF further suppresses the contribution from these low-mass systems.
Finally, while combining $L(M)$ with the CDM HMF reproduces the LF across all luminosities, in WDM scenarios this reconstruction is only possible down to the minimum luminosity associated with the $L(M)$ plateau. Below this point the predicted abundance drops sharply, as shown in Fig.~\ref{fig:reconstructed_LF}, reflecting the underlying paucity of low-mass haloes.

\section{The `observed' \cii power spectrum}\label{sec:CII_power_spectrum}

In this section, we present predictions for the \cii LIM PS that will be observed with the FYST DSS at $z\simeq3.6$ and explore how it is affected by changes in the nature of DM. All technical aspects concerning observational setup, instrumental effects, foreground removal, and PS error estimation follow the methodology outlined in M25; here we only summarise the key elements and refer the reader to that work for further details.
In our baseline configuration, a $16\,\mathrm{deg}^2$ patch of sky is mapped with a total integration time of $8000$ hours, a spectral resolution of $R=100$, and thus a resulting white-noise level of $P_\mathrm{WN} \simeq 2.4\times 10^{10}\,h^{-3}\,\mathrm{Mpc}^3 \,\mathrm{Jy}^2 \,\mathrm{sr}^{-2}$. In the remainder of this paper, we refer to this configuration as `reference setup'.

To investigate the dependence of the results on survey design, we considered variations in sky coverage, sensitivity and spectral resolution ($R$). Specifically, we analysed configurations with areas of $160\,\mathrm{deg}^2$, $1600\,\mathrm{deg}^2$, and half of the sky. In parallel, we examined a scenario where the instrumental sensitivity is enhanced by a factor of $\sqrt{10}$, leading to a tenfold reduction in $\Pwn$.
Finally, we assessed the performance of a futuristic instrument with improved resolution, increasing $R$ to 500 and rescaling the total survey time accordingly.

\begin{figure*}
\centering
\includegraphics[width=0.39\textwidth]{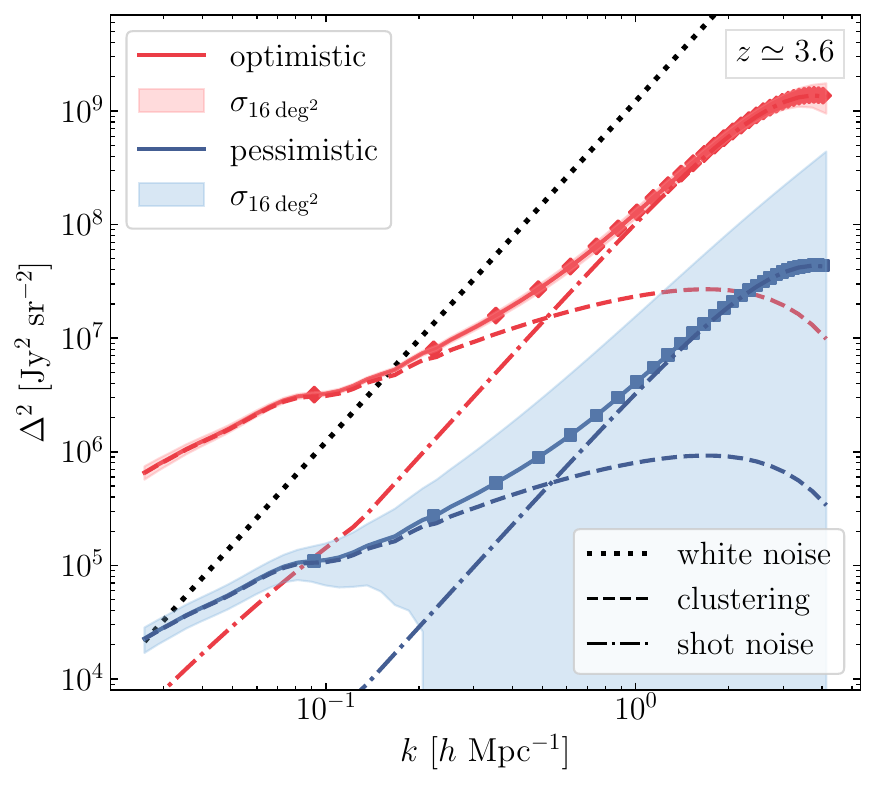}  \hspace{1.5cm}
\includegraphics[width=0.39\textwidth]{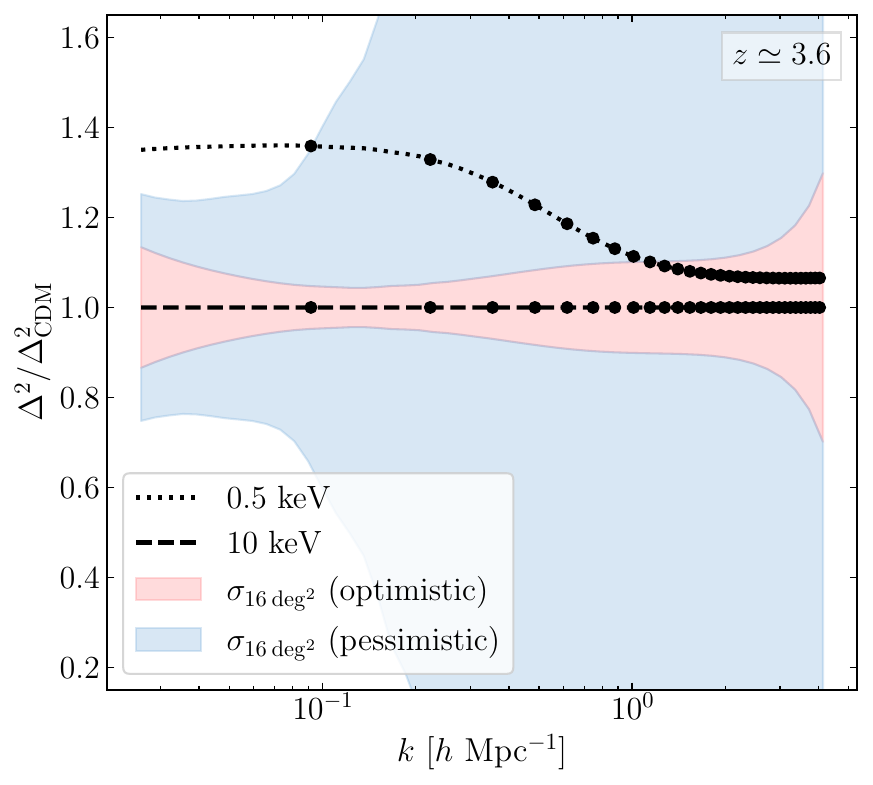} 
\caption{Predicted \cii PS. Left: $\Delta^2(k,z\simeq 3.6)$ computed in the CDM scenario for our optimistic (solid red line) and pessimistic (solid blue line) cases for $\alpha=-1.1$. The shaded areas represent the associated statistical uncertainty as for our survey reference setup. The dotted line is a graphical representation of the white-noise level, $\Pwn$, while the dashed and dot-dashed lines refer to the clustering and shot-noise components, respectively.
Right: Ratio between the WDM and CDM power spectra. The dashed and dotted black lines correspond to WDM models with $m_\mathrm{WDM} = 10$ keV and $0.5$ keV, respectively. The shaded regions reflect the CDM uncertainties from the left panel.
In both panels, the markers illustrate the adopted binning scheme.}
\label{fig:PS_CDM_WDM}
\end{figure*}

To obtain a realistic estimate of the observed \cii PS, we started from its theoretical definition given in Eq.~(\ref{eq:theoretical_LIM_PS}) and related equations, incorporating the instrumental effects through the following expression:
\begin{equation}
    P_\mathrm{obs}(k, \mu, z) = P(k,\mu,z)\, W_\perp(k,\mu)\, W_\parallel(k,\mu)\;,
    \label{eq:Ps_obs}
\end{equation}
where $W_\perp(k,\mu)$ accounts for the Gaussian smoothing due to the telescope beam, and $W_\parallel(k,\mu)$ describes the Lorentzian damping introduced by the finite spectral resolution.
We then defined the direction-averaged PS (i.e. the monopole moment) as
\begin{equation}
    P_0(k,z)=\frac{\int_{\mu_\mathrm{min}}^{\mu_\mathrm{max}} P_\mathrm{obs}(k,\mu,z)\,\mathrm{d}\mu}{\int^{\mu_\mathrm{max}}_{\mu_\mathrm{min}}\mathrm{d}\mu} \; ,
    \label{eq:observedPk_CCAT}
\end{equation}
where $\mu_\mathrm{min} = k_\mathrm{f}^\parallel / k$ and $\mu_\mathrm{max} = \min(1, k_{\mathrm{max}}^\parallel/k)$, with $k_\mathrm{f}^\parallel$ and $k_{\mathrm{max}}^\parallel$ being the fundamental and maximum wavenumbers along the line of sight, respectively. The integration limits were selected from the full range $[0,1]$ to approximately mitigate continuum foregrounds\footnote{Continuum foregrounds primarily affect modes with small line-of-sight components, i.e. $k_\parallel \simeq 0$ \citep[see e.g.][]{Switzer+19, Zhou+23}. As for line interloper contamination, it is expected to be mild above 350 GHz, with fewer than 10\% of voxels affected at 410 GHz \citep[see e.g.][]{Karoumpis+24}; we therefore neglect it in our forecasts.} and to account for the effects of finite spectral resolution.

The associated uncertainty is given by
\begin{equation}
    \sigma_{P_0}(k) = \frac{P_0(k) + \Pwn}{\sqrt{N_\mathrm{m}(k)}}\;,\label{powerspectrum_error}
\end{equation}
where $N_m$ denotes the number of independent Fourier modes contributing to each $k$-bin and can be defined as
\begin{equation}
    N_\mathrm{m}(k) = \frac{\mathrm{min}(k,k_\mathrm{max}^\parallel) \, k \, \Delta k \,  V_\mathrm{surv}}{4 \pi ^2} \; ,
    \label{eq:nmodes}
\end{equation}
with $V_\mathrm{surv}$ being the survey volume.
For practical purposes, we evaluated $P_0$ and $\sigma_{P_0}$ in discrete $k$-bins with fixed width $\Delta k$. Throughout the analysis, we adopted $\Delta k = 5 \, k_\mathrm{f}^\parallel$, but we have verified that our results are not sensitive to this specific choice.

Fig.~\ref{fig:PS_CDM_WDM} (left panel) displays the predicted \cii PS for the reference setup. We plot the quantity $\Delta^2$, which is connected to $P_0$ in Eq.~(\ref{eq:observedPk_CCAT}) via the following relation:
\begin{equation}
    \Delta^2(k,z) = \frac{k^3}{2\pi^2}\,P_0(k,z)\;,
\end{equation}
Solid lines represent the theoretical model, while shaded regions denote the statistical uncertainty derived from Eq.~(\ref{powerspectrum_error}). Red and blue curves correspond to our optimistic and pessimistic LF assumptions, respectively, both adopting a faint-end slope of $\alpha = -1.1$. The discrete points overlaid on each curve indicate the binned measurements resulting from the actual $k$-binning procedure.

The right panel of Fig.~\ref{fig:PS_CDM_WDM} shows the ratio of the WDM and CDM power spectra, focusing on the two most extreme WDM scenarios considered in our Bayesian analysis (see Sect.~\ref{sec:bayesian_inference}). Specifically, these examples were selected to show how models associated with very different WDM particle masses compare to the CDM prediction.
We underline that the higher signal amplitude found in WDM scenarios results from an $L(M)$ relation that exceeds the CDM counterpart at halo masses around the HMF cutoff, where the reduced abundance of WDM haloes is compensated by their higher luminosities. This follows from the requirement that both models reproduce the observed bright end of the \cii LF, thereby raising the WDM luminosities at intermediate halo masses (i.e. $10^{11}-10^{12} \, h^{-1} \, \mathrm{M}_\odot$). Because these masses also provide the bulk of the \cii PS contribution (see Fig.~\ref{fig:weight_masses}), this compensation directly affects the overall amplitude. As already mentioned in the Introduction, an analogous feature has also been reported in \hi studies \citep{Carucci15_WDM_HI}.  This behaviour is further examined in Sect.~\ref{subsec:limitation_weight_masses}, where we discuss the modelling choices and assumptions that lead to this outcome.

\section{Bayesian inference}\label{sec:bayesian_inference}

We investigate the potential of \cii PS measurements to constrain the nature of DM through a Bayesian analysis on mock observations. The focus is on two key parameters: the WDM particle mass, $m_\mathrm{WDM}$, and the line-of-sight displacement, $\sigma$, which partially encapsulates the effect of RSDs.
The statistical framework follows standard practice and employs a Gaussian likelihood function:
\vspace*{0.05cm}
\begin{equation}
    \mathcal{L}(\boldsymbol{\theta}|\mathbf{D}) \propto  \exp\left\{-\frac{1}{2} \sum_i \frac{\left[D_i-M_i(\boldsymbol{\theta}) \right]^2}{\sigma_i^2}\right\} \; ,
    \label{gaussian_likelihood}
\end{equation}
\vspace*{0.05cm}
with $\boldsymbol{\theta}\equiv\{\sigma,m_\mathrm{WDM}\}$ being the model parameters, $\mathbf{D}\equiv\{D_i\}$ the mock data\footnote{In our analysis, producing mock data means treating a halo-model PS prediction computed for a given set of parameters as the underlying truth rather than generating full simulated intensity maps.} (i.e. the \cii PS monopole evaluated in $k_i$-bins), $M_i$ the corresponding model predictions for a given parameter set, and $\sigma_i$ the statistical uncertainties, when the covariance matrix is assumed to be diagonal.

In Section~\ref{subsec:CDM_data}, we assume CDM as the true underlying model and assess which lower bounds on $m_\mathrm{WDM}$ can be placed at a 95\% CL by fitting the mock data with WDM-based models, thus ruling out excessively warm scenarios. In Section~\ref{subsec:3keV_data}, we consider data generated with $m_\mathrm{WDM}=3\,\mathrm{keV}$ and examine whether the true mass can be successfully recovered with both upper and lower limits.

When producing the mock data, a fiducial value of $\sigma = 3\,h^{-1}\,\mathrm{Mpc}$ was adopted to describe the damping caused by RSDs along the line of sight, following the rationale provided in Sect. 4.6 of M25. Throughout the analysis, the LF parameters were assumed to be precisely known from independent observations. Results are presented for both our optimistic and pessimistic LFs with a faint-end slope $\alpha=-1.1$. We assess how the constraints vary under improvements in survey area, sensitivity, and spectral resolution as discussed in Sect.~\ref{sec:CII_power_spectrum}.

\subsection{CDM-generated data}\label{subsec:CDM_data}

In this section, we consider mock data generated under a CDM cosmology, aiming to determine a lower limit for the WDM particle mass across different survey configurations and LF assumptions.

A flat prior was imposed on the displacement parameter $\sigma$ within the interval $[0.2, 5]\,h^{-1}\mathrm{Mpc}$. The prior range for $\sigma$ spans values typical of central galaxies in low-mass haloes up to those representative of satellite systems, including contributions from internal gas dynamics and large-scale relative motions. As our main focus is on $m_\mathrm{WDM}$, we marginalised the posterior distributions over $\sigma$ in order to get the final constraints.
For the WDM particle mass, $m_\mathrm{WDM}$, we considered values above $0.5\,\mathrm{keV}$ because lower masses are strongly disfavoured by existing constraints from galaxy clustering, Lyman-$\alpha$ forest, and reionisation \citep[e.g.][]{Menci+12, Dayal+17,Irsic+17}.

Performing statistical inference on $m_\mathrm{WDM}$ in a CDM universe requires particular care as the CDM case formally corresponds to the limit $m_\mathrm{WDM}\to +\infty$.
In the literature, this is commonly addressed by reparameterising the problem in terms of the inverse particle mass $w=(m_\mathrm{WDM}/\mathrm{keV})^{-1}$ \citep[e.g.][]{Markovic+11, Smith_Markovic_11, Irsic+17, Enzi+21, Rudakovskyi+21, Villasenor+23, Irsic+24}.
A flat prior in $w$ then corresponds to an informative prior in $m_\mathrm{WDM}$ that scales as $\pi_{m_\mathrm{WDM}}\propto m_\mathrm{WDM}^{-2}$, thereby assigning  higher a priori probability to lower particle masses.\footnote{In general, a prior $\pi_w\propto w^\beta$ corresponds to $\pi_{m_\mathrm{WDM}}\propto m_\mathrm{WDM}^\gamma$ with $\gamma=-(\beta+2)$.} 
Since current observations can only provide lower limits on $m_\mathrm{WDM}$, this choice can be regarded as conservative.

Several authors \citep[e.g.][]{Rudakovskyi+21, Villasenor+23}  have examined the dependence of their results on the prior by instead adopting a uniform distribution in $m_\mathrm{WDM}$, truncated at a finite maximum value $m_\mathrm{WDM}^\mathrm{max} \simeq 10 \ \mathrm{keV}$.
This range necessarily excludes the formal CDM limit, as including it would require an improper prior.
In practice, the marginalised likelihood from current datasets is insensitive to larger $m_\mathrm{WDM}$ values, since they yield predictions indistinguishable from CDM within measurement uncertainties.
However, when setting a lower limit on $m_\mathrm{WDM}$ at a fixed CL (e.g. 95\%), the result inevitably depends on $m_\mathrm{WDM}^\mathrm{max}$ through the posterior normalisation.
Consequently, the inferred lower bound can be artificially shifted by changing $m_\mathrm{WDM}^\mathrm{max}$ (see Appendix~\ref{examplebayes} for an illustrative example).

To assess the sensitivity of our analysis to prior assumptions, we explored several alternative forms.
Motivated by the absence of a physically preferred parameterisation, we adopted a family of priors $\pi_w \propto w^{\beta}$ with a uniform hyperprior on $\beta \in [-0.9, 0]$, where the upper limit $\beta = 0$ corresponds to the commonly used flat prior in $w$.
It is worth noting that the case $\beta \to -1$ approaches a flat prior in the logarithm of $m_\mathrm{WDM}/\mathrm{keV}$ ($\gamma = -1$), which cannot be realised exactly because it would be improper; for this reason, we restricted the hyperprior range to values slightly above $-1$.
Over the selected domain $w\in(0,2]$, the normalised prior distribution with a fixed value of $\beta$ can be expressed as
\vspace*{0.05cm}
\begin{equation}
    \pi_w = \frac{1+\beta}{2^{1+\beta}}w^\beta \, .
\end{equation}
\vspace*{0.05cm}
In the remainder of the analysis, all results are presented after marginalising over the hyperprior, which is equivalent to using
\begin{equation}
    \pi_w =\frac{  (y - 1) - (2/w)^{0.9} (-0.9 y + y - 1)}{1.8 y^2}\;,
\end{equation} 
with $y=\ln(w/2)$. This prior probability distribution has its $95^\mathrm{th}$ percentile at $w\simeq 1.821$, corresponding to $m_\mathrm{WDM}\simeq 0.549$ keV.

We also tested a (regularised) two-dimensional Jeffreys prior, proportional to the square root of the determinant of the Fisher information matrix, which ensures reparameterisation invariance but is not necessarily non-informative. In our setup, the Fisher information decreases rapidly for $m_\mathrm{WDM}\gtrsim10\,\mathrm{keV}$, as the predicted power spectra become increasingly insensitive to the particle mass. As a result, the Jeffreys prior suppresses the region corresponding to the CDM limit ($w\to0$) used to generate the mock data, effectively penalising the true model. This suppression shifts the posterior peak away from both the likelihood maximum and the fiducial parameters. Since this shift reflects the structure of the prior rather than information contained in the data, we do not show results obtained with the Jeffreys prior. For completeness, we note that the resulting lower bounds on $m_\mathrm{WDM}$ are weaker by a factor of $1.2$--$1.3$, depending on the configuration considered.

\begin{figure*}[]
\centering
\includegraphics[width=0.3\textwidth, trim=0 0 70 0, clip]{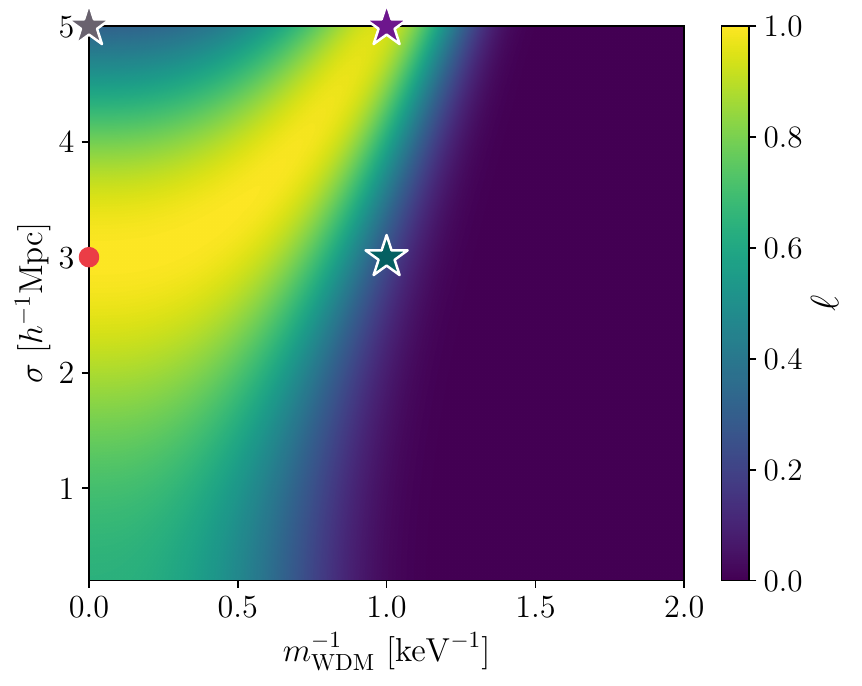}  \hspace{-0.21cm}
\includegraphics[width=0.34\textwidth,trim=25 0 0 0, clip]{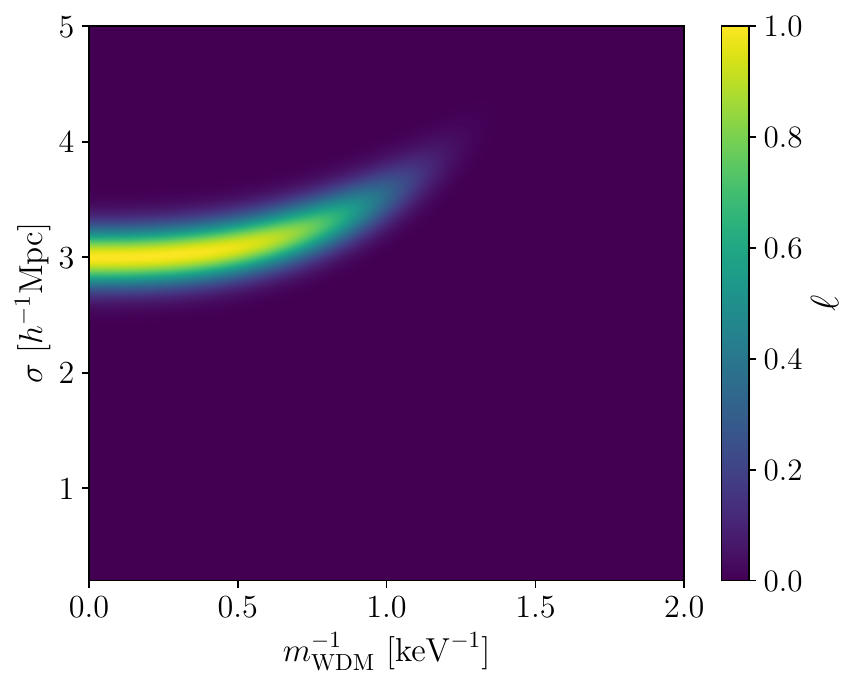}
\includegraphics[width=0.33\textwidth]{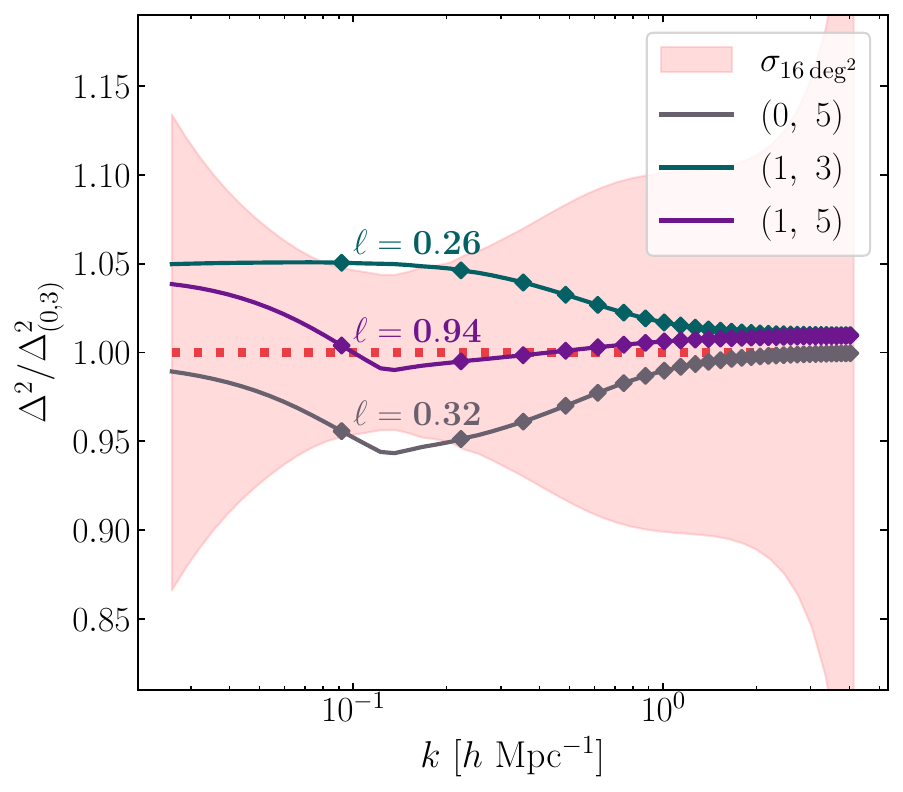} 
\caption{Likelihood ratio $\ell$ computed for the optimistic case. Left: $\ell$ in the reference setup. Centre: Same as the left panel, but with $R=500$. Right: Graphical explanation of the banana-shaped $\ell$ shown in the left panel. Each pair of values $(w,\sigma)$ in the legend indicates the inverse particle mass in keV$^{-1}$ and the RSD displacement parameter in units of $h^{-1} \ \mathrm{Mpc}$, respectively. The stars of corresponding colours in the left panel serve as identifiers for each combination in the parameter space, while the red circle denotes the true values used to generate the data.}
\label{fig:lik2D_optimistic}
\end{figure*}

\begin{figure*}
\centering
\includegraphics[width=0.35\textwidth]{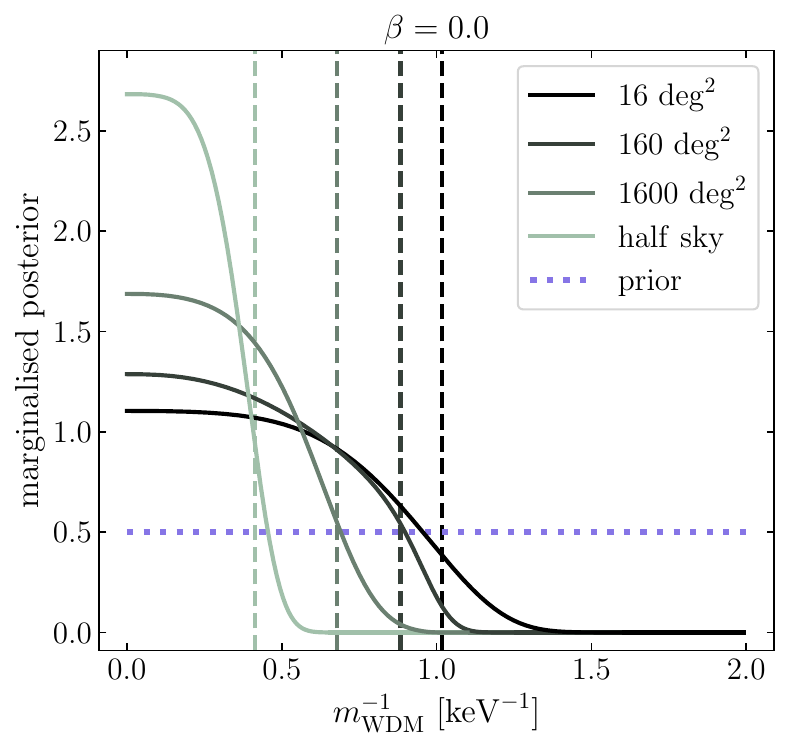} \hspace{1.5cm}\includegraphics[width=0.35\textwidth]{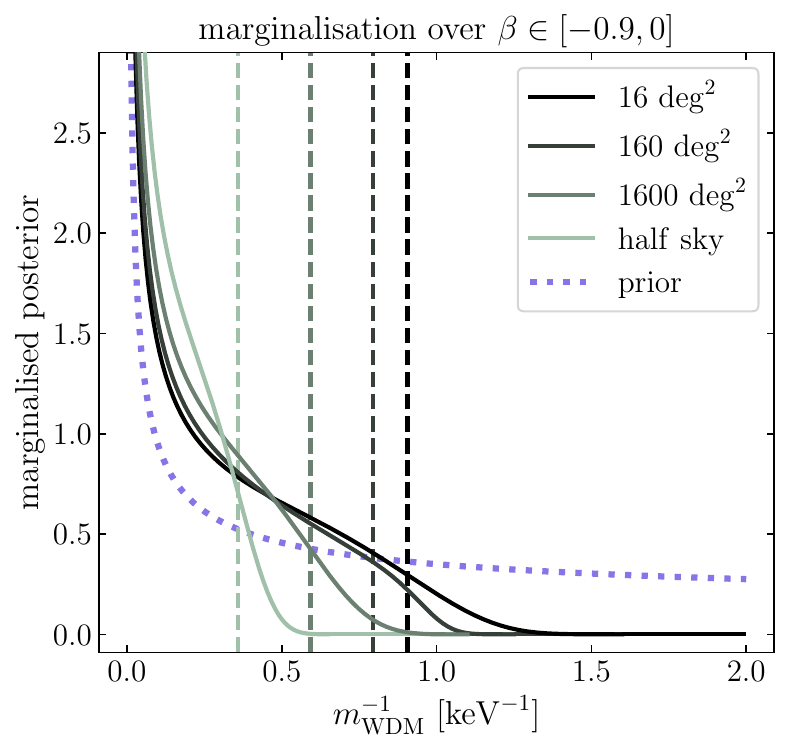} \caption{Marginalised posterior (over $\sigma$) computed for the whole set of sky areas in the optimistic scenario, with $R=100$ and $P_\mathrm{WN} \simeq 2.4\times 10^{10} \,h^{-3}\,\mathrm{Mpc}^3 \,\mathrm{Jy}^2 \,\mathrm{sr}^{-2}$. Left: Posterior for a fixed $\beta=0$ (uniform prior in $w$). Right: Result marginalised over $\beta$. The dashed lines of corresponding colours indicate the $m_\mathrm{WDM}$ threshold up to which CDM and WDM can be distinguished (95\% CL). The dotted lines show the assumed prior distribution for $w$.}
\label{fig:example_marginalised_posterior}
\end{figure*}

As anticipated, we performed the analysis considering both the optimistic and pessimistic LF assumptions, each characterised by a faint-end slope of $\alpha = -1.1$. In all cases, we find that WDM masses above $10\,\mathrm{keV}$ are virtually indistinguishable from CDM at the scales probed by LIM.
The left panel of Fig.~\ref{fig:lik2D_optimistic} illustrates a representative two-dimensional likelihood ratio, $\ell = \mathcal{L}(\boldsymbol{\theta}|\mathbf{D})/\mathcal{L}(\boldsymbol{\theta}_\mathrm{best-fit}|\mathbf{D})$, in the optimistic scenario, assuming the reference survey configuration.
The likelihood contours reveal that WDM models with low $m_\mathrm{WDM}$ values can still provide excellent fits to the CDM mock data, provided the displacement parameter $\sigma$ associated with RSDs is sufficiently large. This reflects a partial degeneracy between the impact of the warm nature of DM and that of the line-of-sight velocity dispersion: Low particle masses enhance power, while large values of $\sigma$ suppress it, allowing the two effects to partially cancel out.
This correlation is exacerbated by the limited spectral resolution, which smooths out small-scale features and makes it difficult to disentangle the scale-dependent damping induced by RSDs from the effects of varying $m_\mathrm{WDM}$.
Indeed, the central panel of Fig.~\ref{fig:lik2D_optimistic} demonstrates how the degeneracy $m_\mathrm{WDM}$--$\sigma$ is mitigated when adopting a higher spectral resolving power, $R = 500$, while keeping the same survey area and instrumental sensitivity. As discussed in M25, increasing $R$ diminishes the effect of spectral smoothing to a level comparable with that of RSDs. This leads to a more accurate estimation of $\sigma$ and significantly weakens its degeneracy with $m_\mathrm{WDM}$.

The compensation mechanism introduced above is further illustrated in the right panel of Fig.~\ref{fig:lik2D_optimistic}, where we compare the fractional difference in the PS signal relative to the fiducial CDM PS (dotted red line) and related uncertainty (shaded region) for three representative cases: (i) a CDM model with $\sigma = 5\,h^{-1}\mathrm{Mpc}$ (solid grey line), corresponding to the upper bound of the prior; (ii) a 1 keV model with $\sigma = 3\,h^{-1}\mathrm{Mpc}$ (solid dark green line), matching the fiducial value for $\sigma$ used to generate the mock data; and (iii) a 1 keV model with $\sigma = 5\,h^{-1}\mathrm{Mpc}$ (solid purple line). The resulting $\ell$ values (0.32, 0.26, and 0.94, respectively) highlight how the interplay between the two parameters of the $m_\mathrm{WDM}$--$\sigma$ plane can yield to good fits even in extreme WDM cosmologies.

Fig.~\ref{fig:example_marginalised_posterior} shows representative examples of the
posterior distribution for $w$, marginalised over $\sigma$. The left panel corresponds to a prior with fixed $\beta=0$, while the right panel uses the effective prior obtained after marginalising over the hyperparameter $\beta$. In both cases, the posterior closely follows the shape of the prior at small values of $w$, where the data provide essentially no constraining power. With $\beta=0$, the posterior approaches a finite plateau as $w \to 0$, whereas the $\beta$-marginalised prior leads to a posterior that diverges in this limit, although it remains normalisable. Importantly, this difference has no practical impact on our analysis: The high-$w$ tail, which drives the 95\% CL, is only marginally affected by the choice of prior. This also confirms that our results do not depend on the arbitrary upper bound of $w=2$ ($m_\mathrm{WDM}=0.5\,\mathrm{keV}$). In this region, the likelihood strongly suppresses low masses, ensuring that the posterior constraints are dominated by the data rather than by prior assumptions.

\begin{figure}
\centering
\includegraphics[width=0.36\textwidth]{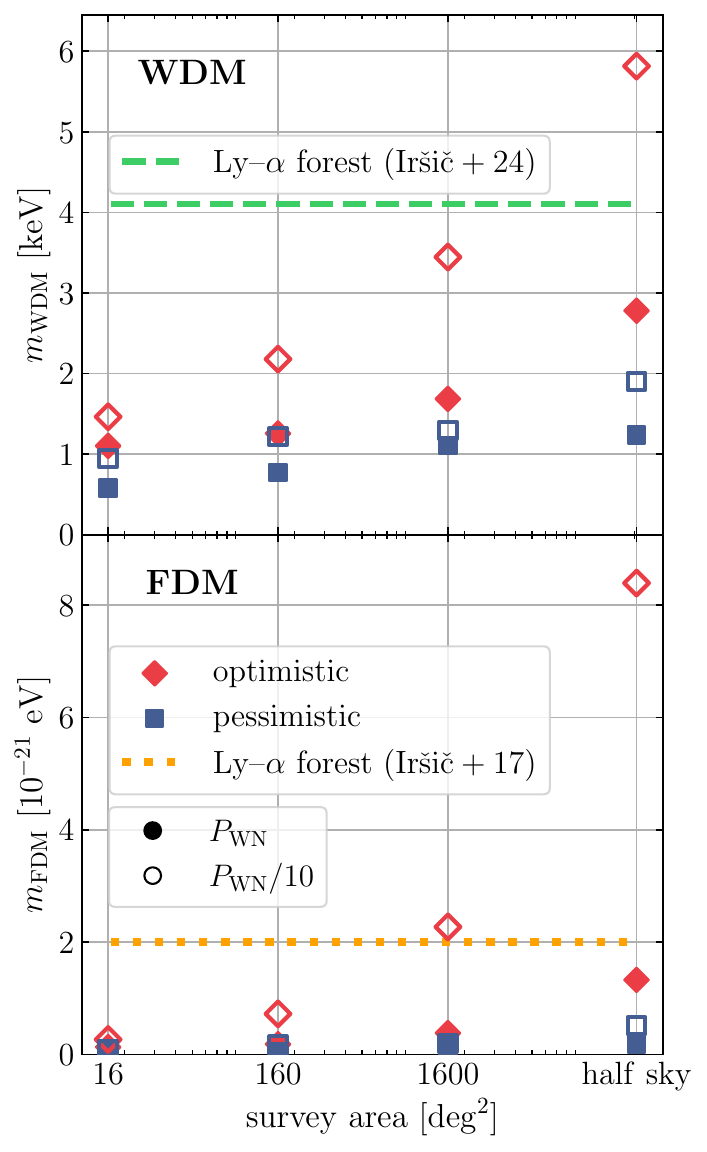} \caption{Lower bounds on $m_\mathrm{WDM}$ (top) and $m_\mathrm{FDM}$ (bottom) for $R=100$. The pessimistic case is represented by the blue squares, while the optimistic case is denoted by the red diamonds. Filled markers correspond to the reference sensitivity; empty markers indicate a sensitivity increased by a factor of $\sqrt{10}$ (i.e. a white-noise PS decreased by a factor of ten). Shown for reference are the dashed green and dotted orange lines, marking the lower bounds inferred by \cite{Irsic+24} and \cite{Irsic+17_fdm}, respectively.}
\label{fig:summary_constraints_R100}
\end{figure}

The resulting constraints, obtained for $R = 100$ and marginalised over $\beta$, for all four considered survey areas, and under both sensitivity settings, are summarised in the top panel of Fig.~\ref{fig:summary_constraints_R100}. The bottom panel displays how our limits on $m_\mathrm{WDM}$ map onto FDM constraints via Eq.~(\ref{eq:FDM_constraints}).
For comparison, Fig.~\ref{fig:summary_constraints_R100} also shows the current limits from Lyman-$\alpha$ forest analyses\footnote{This comparison is not strictly fitting, since our inference was carried out on simulated CDM data, whereas the Lyman-$\alpha$ constraints are based on real observational measurements.} \citep[][]{Irsic+17_fdm,Irsic+24}, which are broadly comparable to those obtained for our $1600 \ \mathrm{deg}^2$ configuration with improved sensitivity. We note, however, that the constraining power of the \cii LIM PS can be further enhanced by incorporating multiple redshift bins and/or additional emission lines (see Sect.~\ref{sec:summary_conclusions}).

\subsection{3 keV WDM-generated data}\label{subsec:3keV_data}

We generated mock data using a WDM model with $m_\mathrm{WDM}=3\,\mathrm{keV}$ to determine whether this particle mass can be reliably recovered within our Bayesian framework, thereby setting both lower and upper bounds. We applied the priors on $\sigma$ and $m_\mathrm{WDM}$ as described in Sect.~\ref{subsec:CDM_data}.

\begin{figure*}
\centering
\includegraphics[width=0.38\textwidth]{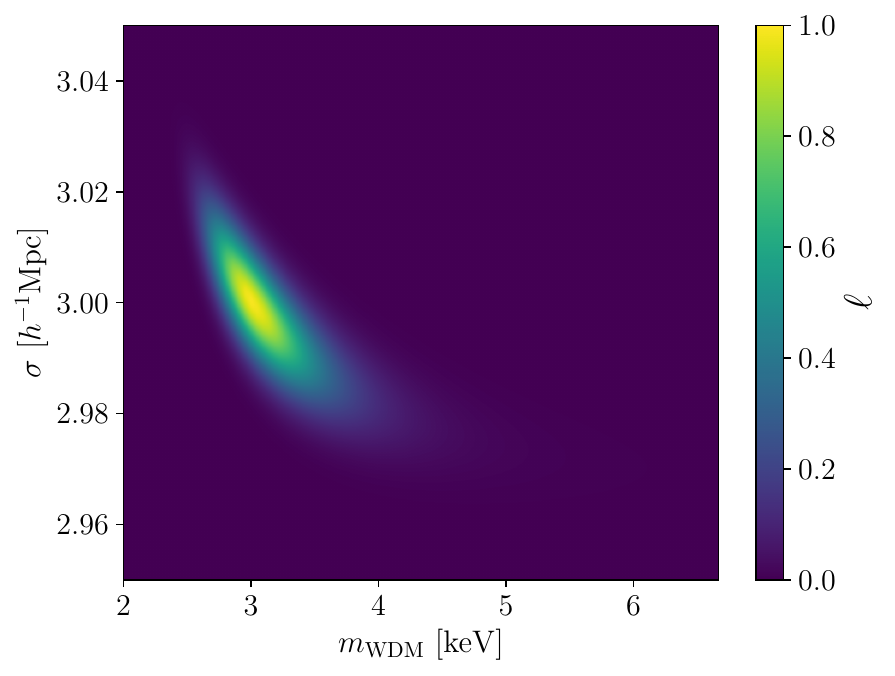}  \hspace{1cm}
\includegraphics[width=0.312\textwidth]{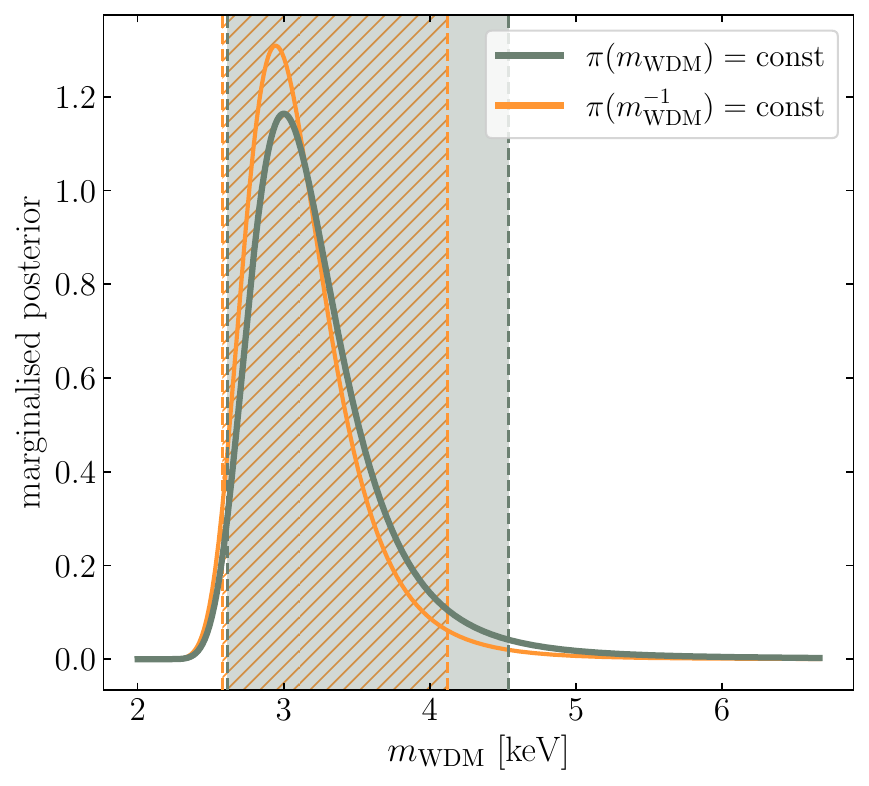}
\caption{Likelihood ratio $\ell$ (left) and marginalised posterior (right) obtained from 3 keV mock data in the optimistic scenario for a survey covering $1600 \ \mathrm{deg}^2$ with $R=500$ and $P_\mathrm{WN} \simeq 2.4\times 10^{9}\,h^{-3}\,\mathrm{Mpc}^3 \,\mathrm{Jy}^2 \,\mathrm{sr}^{-2}$. In the right panel, the dark-green and orange vertical lines denote the 95\% credible regions, computed assuming uniform priors in $m_\mathrm{WDM}$ and $m_\mathrm{WDM}^{-1}$, respectively.}
\label{fig:3keV_2Dlik_posterior}
\end{figure*}

For most survey configurations, $\ell$ remains significantly non-zero even for the CDM scenario ($m_\mathrm{WDM}\to\infty$). As a result, we cannot set meaningful upper limits at the 95\% CL, as CDM remains consistent with the $3\,\mathrm{keV}$ mock data within the errors, in line with the findings discussed in the previous subsection.
Under favourable survey setups, however, the likelihood constrains both parameters tightly, allowing us to determine both upper and lower bounds on $m_\mathrm{WDM}$. An illustrative example is presented in Fig.~\ref{fig:3keV_2Dlik_posterior}, where we show the 2D $\ell$ and the posterior marginalised over $\sigma$ as computed in the optimistic scenario for a survey with $R=500$, $P_\mathrm{WN} \simeq 2.4\times 10^{9}\,h^{-3}\,\mathrm{Mpc}^3 \,\mathrm{Jy}^2 \,\mathrm{sr}^{-2}$, and a sky coverage of $1600~\mathrm{deg}^2$. In this case, uniform priors on $m_\mathrm{WDM}$ can be safely adopted, without any additional treatment of the CDM limit. The resulting posterior and corresponding constraints at 95\% CL (i.e. $2.62<m_\mathrm{WDM}$\! [keV] $<4.54$) are displayed as dark-green lines in the right-hand panel. For comparison, we overplot in orange those obtained by assuming a uniform prior on $w=1/m_\mathrm{WDM}$.
Both approaches produce similar results, although the latter slightly favours lower particle masses as expected from the discussion in Sect.~\ref{subsec:CDM_data}.

\vspace*{0.5cm}

\section{Impact of the faint-end slope}\label{sec:discussion}

In this section, we critically assess the robustness of our results by examining how key modelling choices influence the inferred constraints on $m_\mathrm{WDM}$, focusing on the CDM-generated data. We analyse the relative contribution of different halo mass scales to the LIM components, and investigate the role of the faint-end slope of the LF, exploring how variations in $\alpha$ and correlated Schechter parameters impact the signal and the subsequent statistical inference in the two cosmological scenarios.

\subsection{Differential contributions by halo mass}\label{subsec:limitation_weight_masses}

In Fig.~\ref{fig:weight_masses}, we show the differential contribution of halo masses to $\bar{I}_\nu$, $b$, and $\bar{n}_\mathrm{eff}^{-1}$, in a CDM universe both for the optimistic LF scenario and across the range of fits obtained with $\alpha\in[-1.9,-0.5]$ in the pessimistic case.
The clustering signal is strongly dominated by DM haloes with mass $M\simeq 10^{11}$--$10^{12} \ h^{-1}$ M$_\sun$ and receives non-vanishing (but still minor) contributions from masses below $10^{10} \ h^{-1}$ M$_\sun$ only when $\alpha<-1.4$.
The high-mass cutoff is provided by the HMF while the low-mass one comes from $L(M)$.
The situation is even more extreme for the effective volume per emitter which basically receives no contributions from haloes with $M<10^{11} \  h^{-1}$ M$_\sun$.
It is worth noting that, although Fig.~\ref{fig:weight_masses} refers to the CDM case, the relative weighting of halo masses remains essentially unchanged in WDM scenarios. The only difference arises in more extreme WDM models combined with a steep faint-end slope of the LF, where the scarcity of low-mass haloes leads $L(M)$ to flatten and both the low- and high-mass cutoffs to be determined by the HMF.

As anticipated in Sect.~\ref{sec:CII_power_spectrum}, the enhanced amplitude of the PS in WDM scenarios when $\alpha=-1.1$ originates from this dominance of intermediate-mass haloes in combination with our assumption of a perfectly known LF parameterisation. Specifically, since the luminosity--mass relation in our model (Fig.~\ref{fig:HAM}) was derived via abundance matching (Sect.~\ref{sec:CII_LF_HAM}) applied to the same LF but using either a CDM or WDM HMF, the resulting $L(M)$ curves overlap at high masses but become systematically higher in the WDM case at masses below the scale where the WDM HMF begins to diverge from its CDM counterpart (see Appendix~\ref{app:HMF_bias}). This feature, built into the model by construction, contributes to the boost in power observed for WDM relative to CDM.

All this implies that the large differences in the low-mass end of the HMF between CDM and WDM do not imprint a strong signature in the \cii PS which is most sensitive to substantially higher halo masses. In other words, the constraining power of LIM experiments
on the mass of WDM particles derives from the (smaller) variations of the HMF and linear bias parameter at $M\simeq 10^{11}$--$10^{12}\, h^{-1}$ M$_\sun$.
Consequently, we are unable to fully exploit the potential of \cii as a tracer to determine the nature of DM. As noted in Sect.~\ref{sec:intro}, unlike \hi\!, which has been employed for previous studies involving LIM PS, \cii is present in low-mass haloes that are abundant in CDM but largely absent in WDM. However, the \cii PS is not particularly sensitive to these small haloes. Enhancing their contribution would require an `unconventional' tracer whose emission is stronger in low-mass haloes and weaker in high-mass ones---potentially as a result of physical processes that inhibit its development in more massive environments.

\subsection{Results for a steeper faint-end slope}
\label{subsec:impact_alpha}

The results presented in Sects.~\ref{sec:CII_power_spectrum} and \ref{sec:bayesian_inference} refer to the assumption of $\alpha = -1.1$, and therefore correspond to a scenario in which low-mass haloes ($M < 10^{10}\,h^{-1}\,\mathrm{M}_\odot$) are negligible contributors to the overall signal (see Sect.~\ref{subsec:limitation_weight_masses}). As a consequence, the expected suppression of power in WDM scenarios due to the absence of this small-halo population becomes irrelevant for our purposes.

\begin{figure}
\centering
\includegraphics[width=0.36\textwidth]{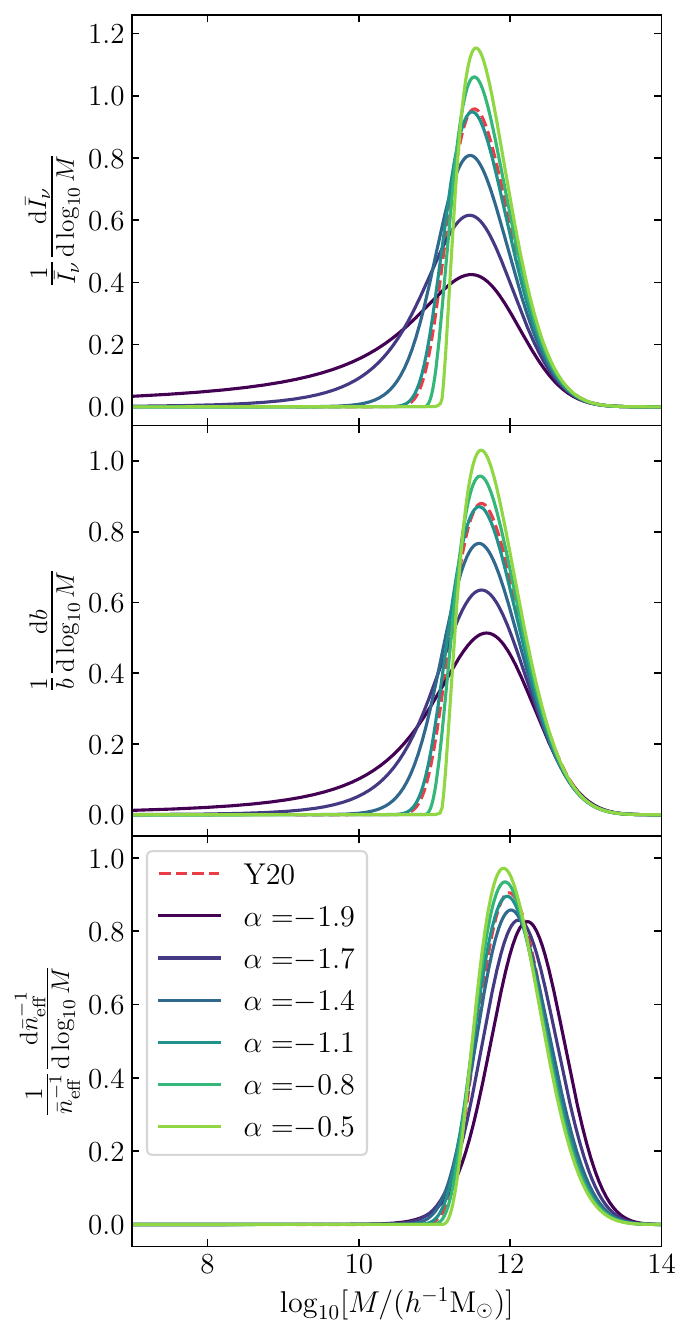} 
\caption{Differential contribution of different halo masses to $\bar{I}_\nu$, $b$, and $\bar{n}_\mathrm{eff}^{-1}$ (from top to bottom) based on our halo model for CDM at $z=3.6$.} 
\label{fig:weight_masses}
\end{figure}

\begin{figure*}
\centering
\includegraphics[width=0.37\textwidth]{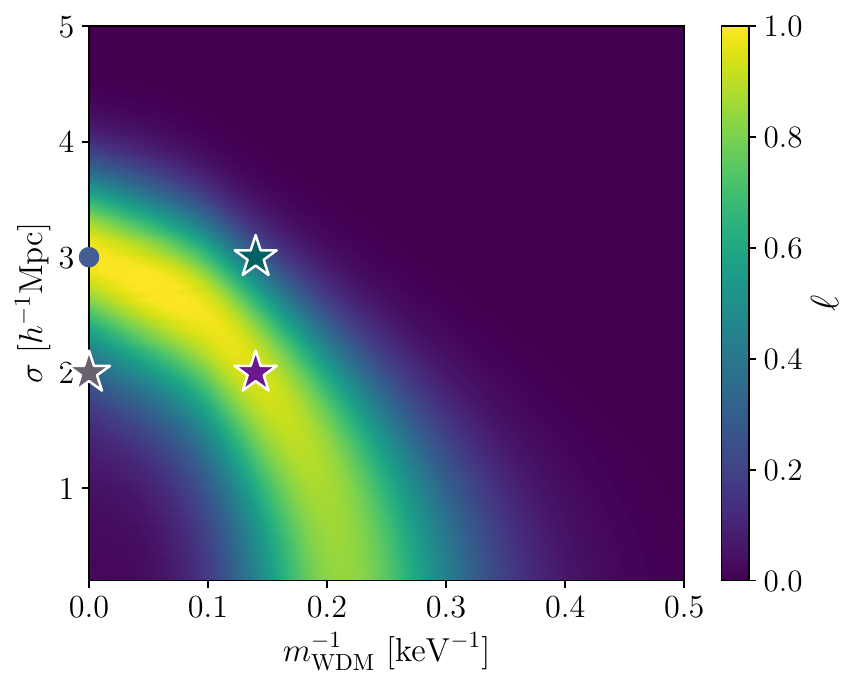} \hspace{1cm}
\includegraphics[width=0.33\textwidth]{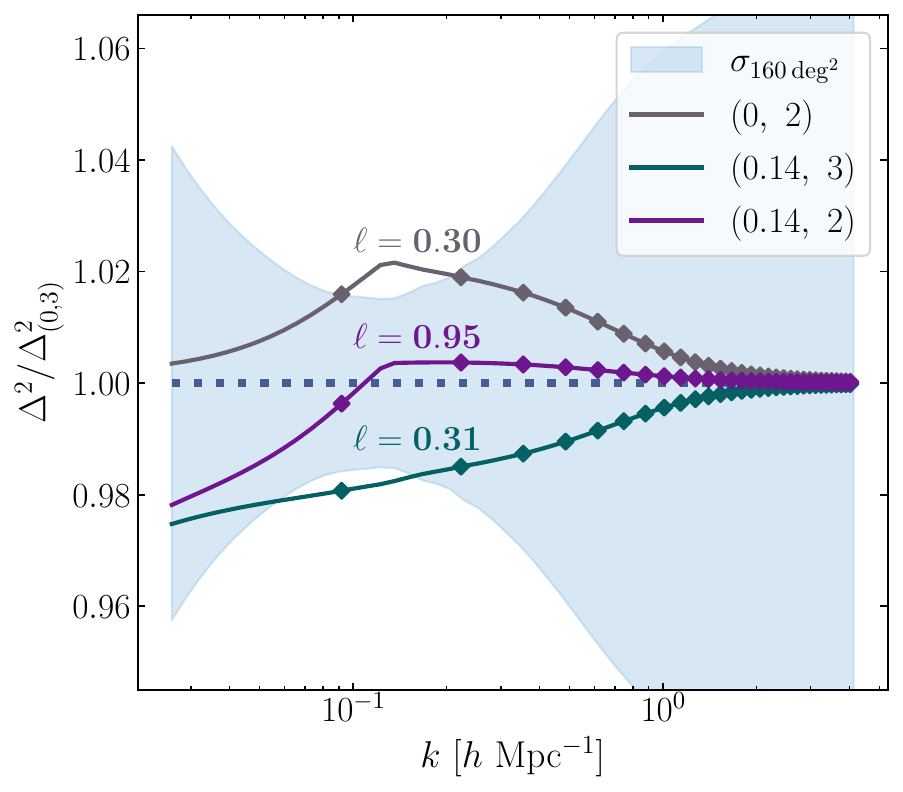} 
\caption{Likelihood ratio $\ell$ computed for the pessimistic case with $\alpha=-1.9$ in a convenient setup ($R=100$, 160 deg$^2$, and $P_\mathrm{WN} \simeq 2.4\times 10^{9}\,h^{-3}\,\mathrm{Mpc}^3 \,\mathrm{Jy}^2 \,\mathrm{sr}^{-2}$) to highlight its reversed shape in the parameter space. The left and right panels are related in the same way as in Fig.~\ref{fig:lik2D_optimistic}.}
\label{fig:lik2D_pessimistic_alpha-1.9}
\end{figure*}

To maximise and exploit the potential contribution of \cii emission from low-mass haloes discussed above, we now turn to steeper faint-end slopes of the LF. In particular, we repeated the analysis for different values of $\alpha$, with special attention to the case $\alpha = -1.9$, which amplifies the role of faint, low-mass haloes in the total emission (see Fig.~\ref{fig:weight_masses}).
In this regime, the degeneracy in the $m_\mathrm{WDM}$--$\sigma$ parameter space discussed in Sect.~\ref{subsec:CDM_data} changes direction, as both parameters now act to suppress the power (left panel of Fig.~\ref{fig:lik2D_pessimistic_alpha-1.9}). Consequently, to reproduce the CDM mock data with $\sigma = 3 \ h^{-1}\ \mathrm{Mpc}$ under increasingly severe WDM scenarios, $\sigma$ must decrease in order to counterbalance the damping induced by RSDs and preserve the overall signal amplitude (right panel of Fig.~\ref{fig:lik2D_pessimistic_alpha-1.9}).

We further find that the improvement in the $m_\mathrm{WDM}$ constraints remains modest relative to the results obtained for the pessimistic LF with $\alpha=-1.1$ in the reference setup, owing to the large uncertainties associated with the mock data. The advantage of a steeper faint-end slope becomes more pronounced when we consider more favourable survey setups.
For example, in a plausible near-future configuration with $R=100$, $160 \ \mathrm{deg}^2$, and $\Pwn\simeq 2.4\times 10^{10}\,h^{-3}\,\mathrm{Mpc}^3 \,\mathrm{Jy}^2 \,\mathrm{sr}^{-2}$, we obtain a 95\% CL of $1.35 \ \mathrm{keV}$. This limit represents an improvement of approximately 75\% relative to the pessimistic case with $\alpha = -1.1$, exceeding even that derived under the optimistic LF assumption of $\sim$8\% for the corresponding survey setup.

\section{Conclusions}\label{sec:summary_conclusions}

The nature of DM remains one of the central open questions in modern cosmology. While the CDM paradigm is the standard framework, alternative scenarios such as WDM composed of thermal relics provide viable solutions to some of the known challenges to $\Lambda$CDM. In this work we have built upon the model of M25, extending it to WDM cosmologies and developing a Bayesian framework to assess the constraining power of the \cii PS on $m_\mathrm{WDM}$. We presented forecasts for the FYST DSS at $z \simeq 3.6$ and explored the potential of future surveys with larger sky coverage, higher sensitivity, and/or increased spectral resolution.
Our analysis led to the following main results:
\begin{itemize}
    \item Under the assumption of a CDM background, the DSS in its reference setup is expected to set a $95\%$ CL lower limit of $1.10$ keV and $0.58$ keV on the WDM particle mass in the optimistic and pessimistic ($\alpha=-1.1$) scenarios, respectively. These bounds improve substantially with more ambitious setups, reaching $5.82$ keV and $1.90$ keV for half-sky coverage and an increase in sensitivity by a factor of $\sqrt{10}$ (see Fig.~\ref{fig:summary_constraints_R100}, with the lower panel showing the corresponding constraints expressed in terms of $m_\mathrm{FDM}$).
    \item Enhancing the spectral resolution from $R=100$ to $R=500$ boosts sensitivity to the damping scale $\sigma$, thereby reducing its degeneracy with $m_\mathrm{WDM}$. In favourable survey conditions and under the optimistic assumption for the LF, this translates into a tightening of constraints by a factor of up to $\sim \ 1.8$. In the reference setup, however, the improvement is negligible, as the likelihood is anyway truncated by the prior imposed on $\sigma$ (Fig.~\ref{fig:lik2D_optimistic}).
    \item By fitting mock data generated with $m_\mathrm{WDM}=3$ keV, we find that the true value can be recovered with both upper and lower bounds only in the optimistic LF scenario and under highly ideal survey configurations. For $R=100$, this requires both half-sky coverage and enhanced sensitivity, while for $R=500$, improved sensitivity combined with a 1600 deg$^2$ footprint is sufficient. In this latter case, for example, we obtain a 95\% CL that lies in $[2.62,4.54]$ keV with uniform priors in $m_\mathrm{WDM}$, and in $[2.58,4.12]$ keV with uniform priors in $m_\mathrm{WDM}^{-1}$, which demonstrates how the latter choice slightly favours the regime of the small WDM particle masses (see Fig.~\ref{fig:3keV_2Dlik_posterior}).
\end{itemize}
In deriving the results presented above, we explored several prior choices: (i) uniform in $m_\mathrm{WDM}$, (ii) uniform in $w=m_\mathrm{WDM}^{-1}$, (iii) proportional to $w^\beta$, marginalising over the hyperparameter $\beta\in[-0.9,0]$, and (iv) a Jeffreys prior.
The latter three yield consistent constraints, whereas the uniform prior in $m_\mathrm{WDM}$ leads to results that depend strongly on the choice of its upper bound (see Appendix~\ref{examplebayes}).

Further insights and additional considerations arising from the analysis are summarised below:
\begin{itemize}
    \item The elongated likelihood contours obtained for $\alpha = -1.1$ (Fig.~\ref{fig:lik2D_optimistic}) arise from the compensating effects of $\sigma$ and $m_\mathrm{WDM}$ on the PS amplitude.
    This degeneracy stems from the form of the $L(M)$ relation derived by matching an extrapolated LF---anchored to existing data but extended to faint, unobserved regimes---to increasingly extreme WDM HMFs, which predict progressively fewer low-mass haloes near and below the cutoff scale (see Sect.~\ref{sec:CII_LF_HAM} and Appendix~\ref{app:HMF_bias}).
    \item As discussed in Sect.~\ref{subsec:limitation_weight_masses}, Fig.~\ref{fig:weight_masses} shows that the LIM signal in terms of the \cii PS is dominated by haloes with masses around $10^{11}$--$10^{12}\,h^{-1}\,\mathrm{M_\odot}$, while low-mass haloes, which mark the sharpest difference between CDM and WDM, contribute only marginally. This limits the sensitivity of current experiments and explains why very ambitious surveys are required to obtain competitive constraints. Nonetheless, adopting a steeper faint-end slope ($\alpha=-1.9$, as discussed in Sect.~\ref{subsec:impact_alpha}) increases the relative contribution of faint sources, reversing the $m_\mathrm{WDM}$--$\sigma$ degeneracy (Fig.~\ref{fig:lik2D_pessimistic_alpha-1.9}) and yielding tighter bounds. The effect is modest in the reference case but grows with survey improvements.  
\end{itemize}
In summary, our forecasts underline both the potential and the current limitations of \cii LIM as a probe of DM physics. While the DSS alone is unlikely to deliver stringent constraints, future surveys combining wider sky coverage, greater sensitivity, and higher resolution will significantly enhance its constraining power.

In this context, it is important to note that this work focuses on a single line and a narrow redshift range\footnote{The DSS also observes at frequencies corresponding to higher redshifts, but the resulting PS measurements are expected to have a much lower signal-to-noise ratio and to be more strongly affected by interloper contamination.} and therefore covers only a limited portion of the available observational domain.
The resulting constraints should be regarded as somewhat conservative, as they do not yet exploit the full potential of forthcoming LIM surveys.
Future LIM experiments will naturally provide access to multiple emission lines and extend coverage across a broad range of redshifts.
Combining information from different epochs enhances the sensitivity to departures from CDM and simultaneously enables a more detailed characterisation of the evolution of the \cii luminosity--mass relation.
Moreover, by jointly analysing signals from multiple tracers, such as \cii\!\!, CO, Ly-$\alpha$, or H$\alpha$, and combining measurements across several redshift intervals, it becomes possible to probe a wider range of halo masses and astrophysical environments, thereby obtaining tighter constraints on $m_\mathrm{WDM}$.
Although the exact quantitative gain depends on the degree of correlation between lines and the overlap in redshift coverage, multi-line and multi-epoch analyses are expected to substantially reduce uncertainties and yield more robust constraints on the nature of DM.
These considerations highlight the strong potential of coordinated LIM surveys to overcome current limitations and maximise their cosmological return.

\begin{acknowledgements}
The authors gratefully acknowledge the Collaborative Research Center 1601 (SFB 1601 sub-project C6) funded by the Deutsche Forschungsgemeinschaft (DFG, German Research Foundation) -- 500700252. They also acknowledge the International Max Planck Research School for Astronomy and Astrophysics (IMPRS A\&A) at the Universities of Bonn and Cologne for supporting EM through a research contract.
EM is a member of the IMPRS A\&A, the Bonn Cologne Graduate School (BCGS), and guest researcher at the Max Planck Institute for Radio Astronomy (MPIfR) in Bonn.
PK held the same affiliations (IMPRS A\&A, BCGS, and MPIfR) at the time when most of this research was carried out.
CP is grateful to SISSA, the University of Trieste, and IFPU, where part of this work was carried out, for hospitality and support.
MV is partially supported by the Fondazione ICSC, Spoke 3 ``Astrophysics and Cosmos Observations'', Piano Nazionale di Ripresa e Resilienza Project ID CN00000013 ``Italian Research Center on High-Performance Computing, Big Data and Quantum Computing'' funded by MUR Missione 4 Componente 2 Investimento 1.4: Potenziamento strutture di ricerca e creazione di ``campioni nazionali di R\&S (M4C2-19)'' - Next Generation EU (NGEU). 
MV is also supported by the INAF Theory Grant ``Cosmological Investigation of the Cosmic Web'' and by the INFN INDARK grant. The research of AMD is supported by the Agence Nationale de la Recherche (ANR), grant ANR-23-CPJ1-0160-01. 

\end{acknowledgements}

\bibliographystyle{aa}
\bibliography{biblio}

\clearpage

\onecolumn

\begin{appendix}

\section{Halo abundance and clustering in CDM and WDM models }\label{app:HMF_bias}
\begin{figure*}[h!]
\centering
\includegraphics[width=0.3\textwidth]{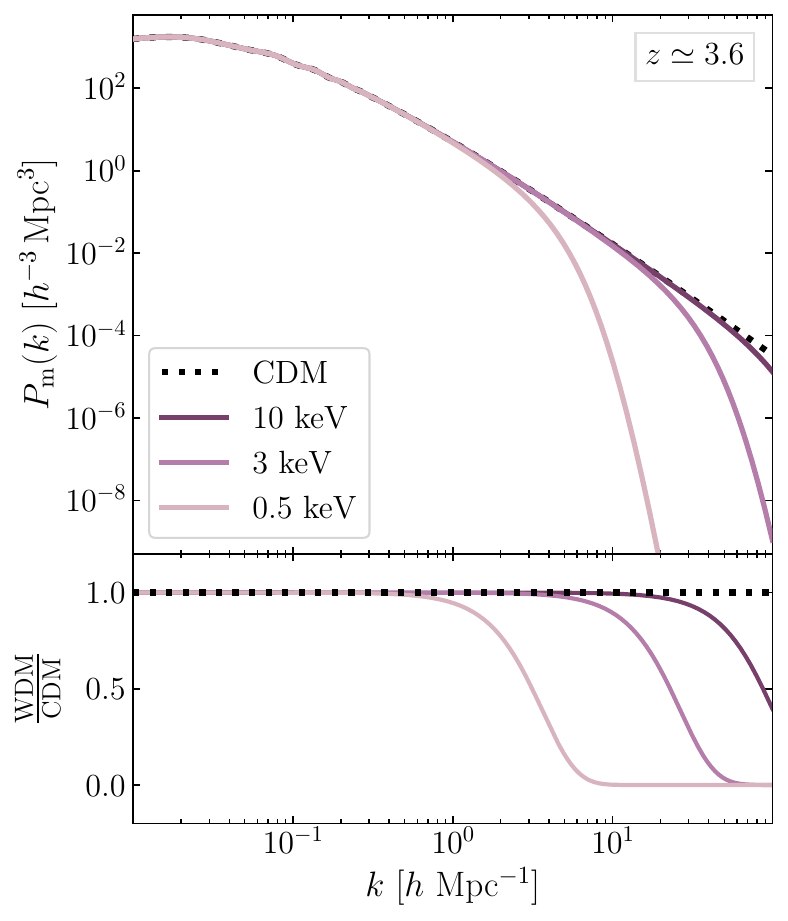} \hspace{0.3cm}
\includegraphics[width=0.311\textwidth]{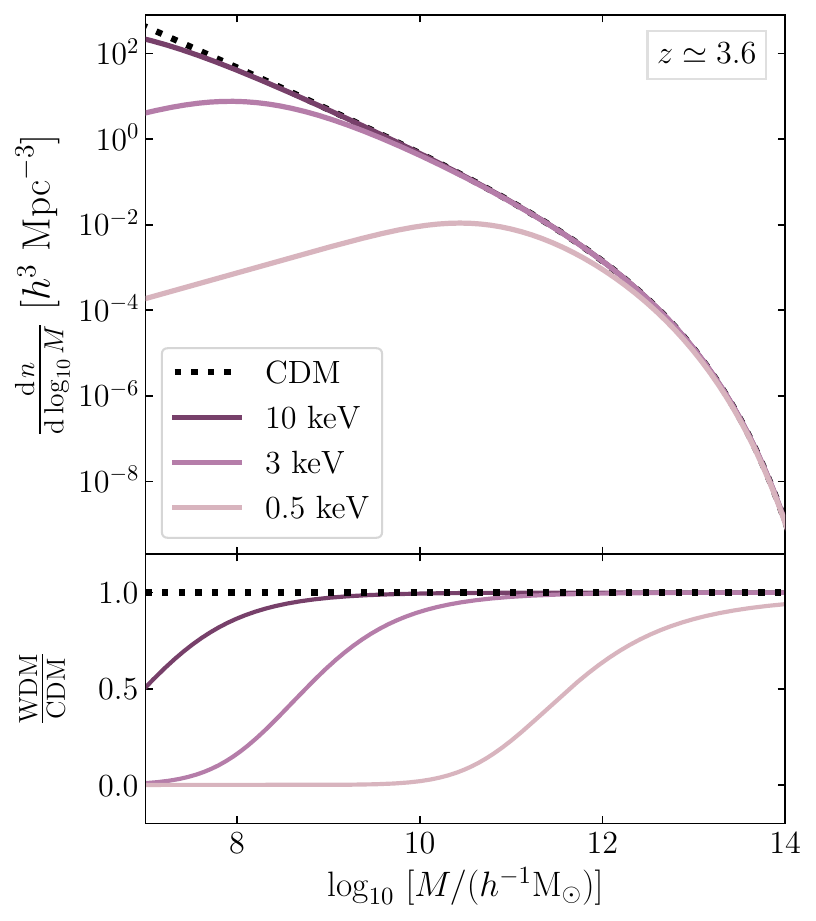} \hspace{0.3cm}
\includegraphics[width=0.302\textwidth]{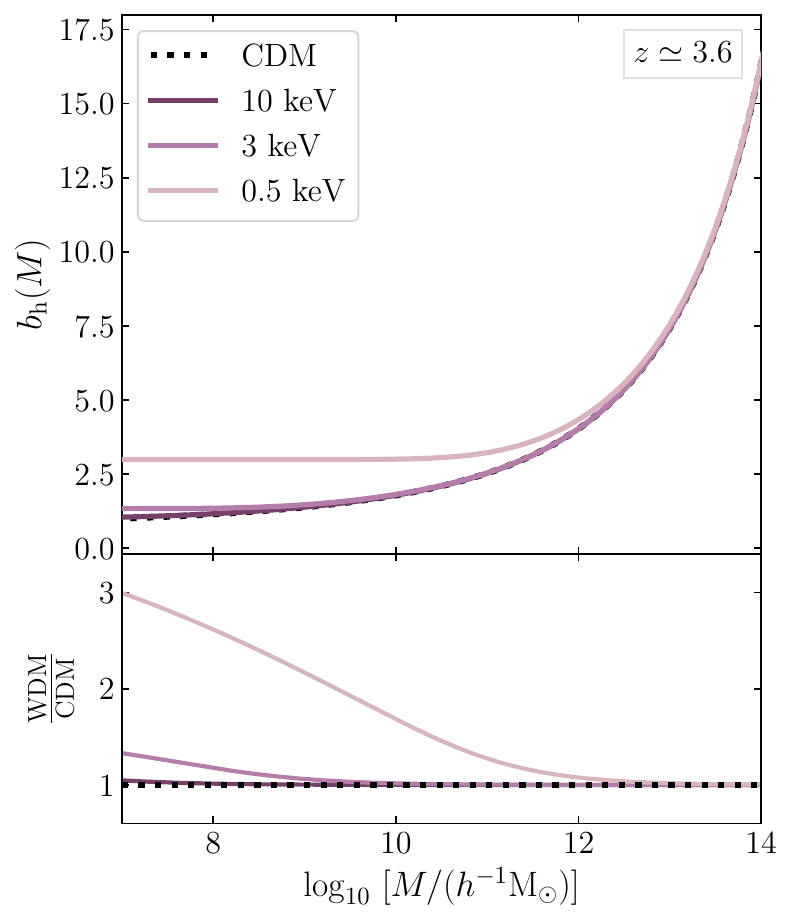} 
\caption{Comparison between CDM and WDM models at $z\simeq3.6$. Top panels: matter PS (left), HMF (centre), halo bias (right). The dotted black line corresponds to CDM, while the coloured lines represent WDM models with three representative particle masses. Bottom panels: Ratio of WDM to CDM, illustrating the deviation of each WDM model from the CDM baseline.}

\label{fig:MF_BIAS_CDM_WDM}
\end{figure*}

We compare CDM and WDM models at $z \simeq 3.6$, highlighting their effects on the matter PS, HMF, and linear bias parameter (Fig.~\ref{fig:MF_BIAS_CDM_WDM}).

\section{Uniform prior in $m_\mathrm{WDM}$}
\label{examplebayes}
For illustrative purposes,
in this Appendix we combine the likelihood function obtained for the optimistic LF in our reference survey setup (see the left panel of Fig.~\ref{fig:lik2D_optimistic}) with
a uniform prior in $m_\mathrm{WDM}$ defined over two different ranges. Because the likelihood becomes flat at large $m_\mathrm{WDM}$, the inferred upper bound 
on the WDM particle mass depends strongly on the chosen
maximum value $m_\mathrm{WDM}^\mathrm{max}$, thereby highlighting the prior dependence of this inference problem.

\begin{figure}[h!]
\centering
\includegraphics[width=0.38\textwidth]{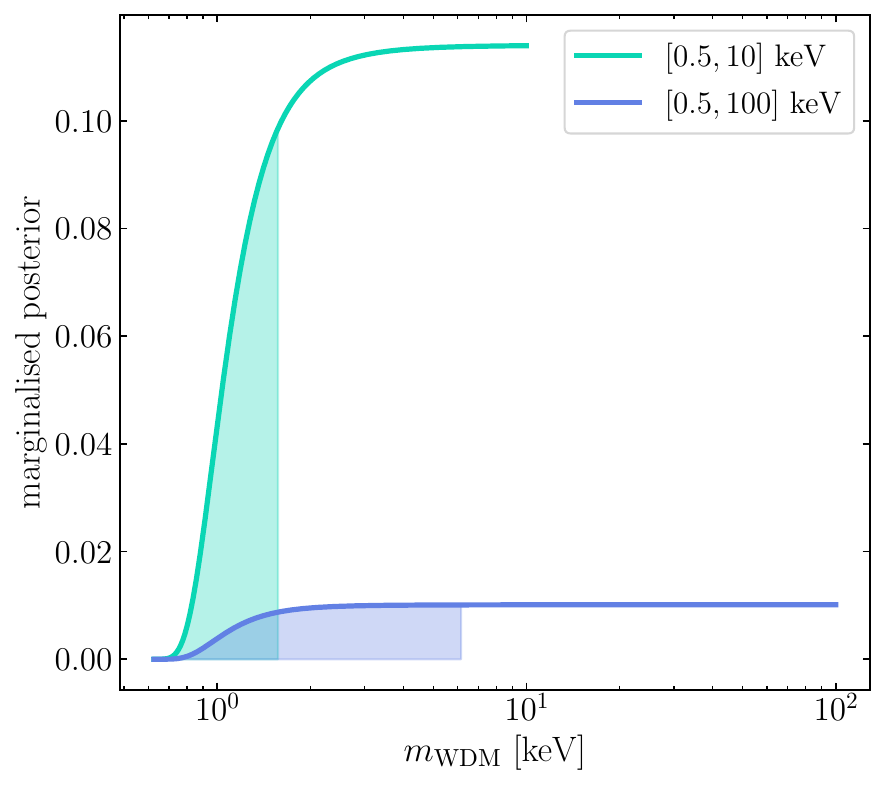} 
\caption{Marginalised posterior distributions (differential in $m_\mathrm{WDM}$) for the reference survey in the optimistic scenario. The posteriors are obtained assuming uniform priors in $m_\mathrm{WDM}$ over two different mass ranges. Shaded regions indicate the 5\% tails, defining the 95\% CL for $m_\mathrm{WDM}$, corresponding to 1.57 keV and 6.16 keV for $m_\mathrm{WDM}^\mathrm{max} = 10$ and $100$ keV, respectively.}
\label{fig:uniform_prior_mWDM_problem}
\end{figure}

\end{appendix}

\end{document}